\documentstyle[12pt,epsf]{article}

\begin{document}

\thispagestyle{empty}

\centerline{{\Large \bf ERMAKOV-LEWIS DYNAMIC INVARIANTS}}

\bigskip
\bigskip

\centerline{{\Large \bf WITH SOME APPLICATIONS}}

\bigskip
\bigskip
\bigskip
\bigskip
\bigskip
\bigskip
\bigskip
\bigskip
\bigskip

\centerline{{\Large Pedro Basilio Espinoza Padilla}}

\bigskip
\bigskip
\bigskip
\bigskip
\bigskip
\bigskip

\centerline{{\Large Master Thesis}}

\bigskip
\bigskip
\bigskip
\bigskip
\bigskip

\centerline{{\Large INSTITUTO DE F\'ISICA}}

\bigskip
\bigskip

\centerline{{\Large UNIVERSIDAD DE GUANAJUATO}}

\bigskip
\bigskip
\bigskip
\bigskip
\bigskip
\bigskip
\bigskip

\centerline{\Large Dr. Haret C. Rosu}
\centerline{Supervisor}

\bigskip
\bigskip
\bigskip

\centerline{Le\'on, Guanajuato}

\bigskip
\bigskip

\centerline{31 January 2000}

\newpage


{\Large \bf Contents.}\hfill \dots 2

\bigskip
\bigskip

1. Introduction.\hfill \dots 3

\bigskip
\bigskip

2. The method of Ermakov.\hfill \dots 4

\bigskip
\bigskip

3. The method of Milne. \hfill \dots 7

\bigskip
\bigskip

4. Pinney's result. \hfill \dots 8

\bigskip
\bigskip

5. Lewis' results. \hfill \dots 8

\bigskip
\bigskip

6. The interpretation of Eliezer and Gray. \hfill \dots 14

\bigskip
\bigskip

7. The connection of the Ermakov invariant with N\"other's theorem. 
\hfill\dots 17

\bigskip
\bigskip

8. Possible generalizations of Ermakov's method. \hfill \dots 20

\bigskip
\bigskip

9. Geometrical angles and phases in Ermakov's problem. \hfill\dots 22

\bigskip
\bigskip

10. Application to the minisuperspace cosmology. \hfill \dots 26

\bigskip
\bigskip

11. Application to physical optics. \hfill \dots 42

\bigskip
\bigskip

12. Conclusions. \hfill \dots 47

\bigskip
\bigskip

Appendix A: Calculation of the integral of $I$. \hfill \dots 48

\bigskip
\bigskip

Appendix B: Calculation of $\langle\hat{H}\rangle$ in eigenstates
of $\hat{I}$. \hfill \dots 49

\bigskip
\bigskip

 References. \hfill \dots 50

\bigskip
\bigskip

\newpage

\section*{1. Introduction.}

In this work we present a study of the   
Ermakov-Lewis invariants that are related to some
linear differential equations of second order and one variable which are of much
interest in many areas of physics. In particular we shall study
in some detail the application of the Ermakov-Lewis formalism
to several simple Hamiltonian models of ``quantum" cosmology. There is also
a formal application to the physical optics of waveguides. 

In 1880, Ermakov \cite{E} published excerpts of his course 
on mathematical analysis where he described a relationship between linear 
differential equations of second order and a particular type of nonlinear 
equation. At the beginning of the thirties, 
Milne \cite{M} developed a method quite similar to the WKB technique where the 
same nonlinear equation found by Ermakov occurred,
and applied it successfully to several model problems in quantum mechanics. 
Further, in 1950, the solution to this nonlinear differential equation has been
given by Pinney \cite{P}. 

On the other hand, within the study of the adiabatic invariants 
at the end of the fifties, a number of powerful perturbative methods in the 
phase space have been developed .
In particular, 
Kruskal \cite{K} introduced a certain type of canonical variables
which had the merit of considerably simplifying the mathematical approach and 
of clarifying some quasi-invariant structures of the phase space. 
Kruskal's results have been used by 
Lewis \cite{L,L68} to prove that the adiabatic invariant found by Kruskal
is in fact a true invariant. Lewis applied it to the well-known  
problem of the harmonic oscillator of time-dependent frequency. 
Moreover, Lewis and Riesenfeld \cite{LR} proceeded to quantize the  invariant, 
although the physical interpretation was still not clear even at the classical 
level. In other words, a constant of motion without meaning was available. 

In a subsequent work of Eliezer and Gray \cite{EG}, an elementary physical 
interpretation was achieved in terms of the angular momentum
of an auxiliary two-dimensional motion. Even though this interpretation
is not fully satisfactory in the general case, it is the clearest at the moment. 

Presently, the Ermakov-Lewis dynamical invariants are more and more
in use for many different time-dependent problems whose 
Hamiltonian is a quadratic form in the canonical coordinates. 

\section*{2. The method of Ermakov.}

\setcounter{equation}{0} 

The Ukrainian mathematician
V. Ermakov was the first to notice that some nonlinear differential equations 
are related in a simple and definite way with the second order linear
differential equations. Ermakov gave as an example the so-called Ermakov system
for which he formulated the following theorem.

\bigskip

{\bf Theorem 1E}.
{\footnotesize {\em If an integral of the equation  
\setcounter{equation}{0}
\begin{equation}
\frac{d^2y}{dx^2}=My
\end{equation}
is known, we can find an integral of the equation 
\begin{equation}
\frac{d^2z}{dx^2}=Mz+\frac{\alpha}{z^3},
\end{equation}
where $\alpha$ is some constant.}} 

\bigskip

Eliminating $M$ from these equations
one gets
\[\frac{d}{dx}\left(y\frac{dz}{dx}-z\frac{dy}{dx}\right)=\frac{\alpha y}{z^3}.\]
Multiplying both sides by 
\[ 2\left(y\frac{dz}{dx}-z\frac{dy}{dx}\right),\]
the last equation turns into
\[\frac{d}{dx}\left(y\frac{dz}{dx}-z\frac{dy}{dx}\right)^2=
-\frac{2\alpha y}{z}\frac{d}{dx}\left(\frac{y}{z}\right).\]
Multiplying now by $dx$ and integrating both sides we get
\begin{equation}
\left(y\frac{dz}{dx}-z\frac{dy}{dx}\right)^2=C-\frac{\alpha y^2}{z^2}.
\end{equation}
If $y_1$ and $y_2$ are two particular solutions of the equation (1), 
substituting them by $y$ in the latter equation we get two integrals of
of the equation (2)
\[\left(y_1\frac{dz}{dx}-z\frac{dy_1}{dx}\right)^2=C_1-\frac{\alpha y_1^2}{z^2},\]
\[\left(y_2\frac{dz}{dx}-z\frac{dy_2}{dx}\right)^2=C_2-\frac{\alpha y_2^2}{z^2}.\]
Eliminating $dz/dx$ from these equations we get a general first integral 
of (2).
One should note that the Ermakov system coincide with the problem of the 
two-dimensional parametric oscillator 
(as we shall see in chapter 6). Moreover, the proof of the theorem gives an exact
method to solve this important dynamical problem.

The general first integral of equation (2) can be also obtained as follows.
Getting $dx$ from equation (3): 
\[dx=\frac{ydz-zdy}{ \sqrt{C-\alpha y^2/z^2} }.\]
Dividing both sides by $y^2$ we get the form
\[\frac{dx}{y^2}=\frac{\frac{z}{y}d\left(\frac{z}{y}\right)}{\sqrt{C\frac{z^2}
{y^2}-\alpha}}.\]
Multiplying by C and integrating both sides we get:
\[C\int\frac{dx}{y^2}+C_3=\sqrt{C\frac{z^2}{y^2}-\alpha}~.\]
This is the general first integral of equation (2), where $C_3$ is the 
constant of the last integration. 
For $y$ is enough to take any particular integral of equation (1).

As a corollary of the previous theorem we can say that

\bigskip

{\bf Corollary 1Ec}. {\footnotesize {\em If a particular solution  
of (2) is known, we can find the general solution of equation (1).}}

\bigskip
 
Since it is sufficient to find particular solutions of (1), we can take $C=0$ 
in equation (3). Thus we get:
\[y\frac{dz}{dx}-z\frac{dy}{dx}=\mp\frac{y}{z}\sqrt{-\alpha}.\]
and therefore
\[\frac{dy}{y}=\frac{dz}{z}\pm\frac{dx\sqrt{-\alpha}}{z^2}.\]
Integrating both sides 
\[\log y=\log z\pm\sqrt{-\alpha}\int\frac{dx}{z^2},\]
which results in
\[y=z\exp(\pm\sqrt{-\alpha}\int \frac{dx}{z^2}).\]
Taking the plus sign first and the minus sign next we get two particular 
solutions of equation (1).

A generalization of the theorem has been given by the same Ermakov.

\bigskip

{\bf Theorem 2E}.
{\footnotesize {\em If $p$ is some known function of $x$ and $f$ is any other arbitrary given
function, then the general solution of the equation 
\[p\frac{d^2y}{dx^2}-y\frac{d^2p}{dx^2}=\frac{1}{p^2}f\left(\frac{y}{p}\right)\]
can be found by quadratures.}}

\bigskip

Multiplying the equation by 
\[2\left(p\frac{dy}{dx}-y\frac{dp}{dx}\right)dx\]
one gets the following form
\[d\left(p\frac{dy}{dx}-y\frac{dp}{dx}\right)^2=2f\left(\frac{y}{p}\right)
d\left(\frac{y}{p}\right).\]
Integrating both sides and defining for simplicity reasons
\[2\int f(z)dz=\varphi(z),\] we get
\[\left(p\frac{dy}{dx}-y\frac{dp}{dx}\right)^2=\varphi\left(\frac{y}{p}\right)+C.\]
This is the expression for a first integral of the equation. Thus, for
$dx$ we have:
\[dx=\frac{pdy-ydp}{\sqrt{\varphi\left(\frac{y}{p}\right)+ C}}.\]
Dividing by $p^2$ and integrating both sides we find:
\[\int\frac{dx}{p^2}+C_4=\int\frac{d\left( \frac{y}{p}\right) }{\sqrt{
\varphi\left(\frac{y}{p}\right)+C} }. \]
This is the general integral of the equation.

A particular case is when $p=x$. Then, the differential equation
can be written as
\[x^3\frac{d^2y}{dx^2}=f\left(\frac{y}{x}\right).\]

\setcounter{equation}{0} 
\section*{3. The method of Milne.}

In 1930, Milne \cite{M} introduced a method to solve the  
Schr\"odinger equation taking into account the basic oscillatory structure of 
the wave function. This method has been one of the first in the class of the 
so-called phase-amplitude procedures, which allow to get sufficiently exact
solutions for the one-dimensional 
Schr\"odinger equation at {\em any} energy and are used to locate resonances.

Let us consider the one-dimensional Schr\"odinger equation
\begin{equation} \label{m1}
\frac{d^2 \psi}{dx^2}+k^2(x)\psi=0
\end{equation}
where $k^2(x)$ is the local wave vector
\begin{equation} \label{m2}
k^2(x)=2\mu[E-V(x)]~.
\end{equation}
Milne proposed to write the wave function as a variable amplitude multiplied
by the sinus of a variable phase, i.e.,
\begin{equation} \label{m3}
\psi(x)=\left(\frac{2\mu}{\pi}\right)^{1/2}\beta(x)\sin(\phi(x)+\gamma)
\end{equation}
where $\mu$ is the mass parameter of the problem at hand, $E$ is the 
total energy of the system, $\gamma$ is a constant phase, 
and $V(x)$ is the potential energy. 
In the original method, 
$\beta$ and $\phi$ are real and $\beta>0$. Substituting the previous 
expression for $\psi$ in the wave equation and solving for $d\phi/dx$
one gets
\begin{equation} \label{m4}
\frac{d^2 \beta}{dx^2}+k^2(x)\beta=\frac{1}{\beta ^3}~,
\end{equation}
\begin{equation} \label{m5}
\frac{d\phi}{dx}=\frac{1}{\beta ^2}~.
\end{equation}
As one can see, the equation for Milne's amplitude coincides with the nonlineal 
equation found by Ermakov, now known as Pinney's equation.

\setcounter{equation}{0} 
\section*{4. Pinney's result.} 

In a brief note Pinney was the first to give without proof 
(claimed to be trivial) the connection between the solutions of the equation for
the parametric oscillator and the nonlinear equation (known as Pinney's 
equation or Pinney-Milne equation). 
\begin{equation}
y''(x)+p(x)y(x)+\frac{C}{y^3}(x)=0
\end{equation}
for $C=$ constant and $p(x)$ given. The general solution for which 
$y(x_{\rm 0})=
y_{\rm 0}$, $y'(x_{\rm 0})=y'_{\rm 0}$ is
\begin{equation}
y_{P}(x)=\left[ {U^2}(x)-C{W^{-2}}{V^2}(x)\right]^{1/2}~,
\end{equation}
where $U$ and $V$ are solutions of the linear equation
\begin{equation}
y''(x)+p(x)y(x)=0~,
\end{equation}
for which $U(x_{\rm 0})=y_{\rm 0}$, $U'(x_{\rm 0})=y'_{\rm 0}$; $V(x_{\rm 0})
=0$, $V'(x_{\rm 0})\neq y'_{\rm 0}$ and $W$ is the Wronskian $W=UV'-U'V=$
constant $\neq 0$ and one takes the square root in (2) for that solution
which at $x_{\rm 0}$ has the value $y_{\rm 0}$

The proof is very simple as follows.
From eq.~(2) we get $\dot{y_{P}}=y_{P}^{-1}(U\dot{U}-CW^{-2}V\dot{V})$ and
$\ddot{y_{P}}=-y_{P}^{-3}(U\dot{U}-CW^{-2}V\dot{V})^{2} + y_{P}^{-1}
(\dot{U}^{2}-CW^{-2}\dot{V}^{2})-p(x)y_{P}$. From here,
explicitly calculating $\ddot{y_{P}}+p(x)y_{P}$ 
one gets $-Cy_{P}^{-3}$ and therefore Pinney's equation.

\setcounter{equation}{0} 
\section*{5. Lewis' results.}

In 1967 Lewis considered parametric Hamiltonians of the standard form:
\begin{equation} \label{l1}
H_{L}=(1/2\epsilon)[p^2+\Omega ^2(t)q^2]~,
\end{equation}

If $\Omega$ is real, the motion of the classical system whoose 
Hamiltonian is given by eq.~(\ref{l1}) is oscillatory with an arbitrary high
frequency when $\epsilon$ goes to zero.
Corresponding to this, there are asymptotic series in positive powers of
$\epsilon$, whose partial sums are the adiabatic invariants of the system;
the leading term of the series is 
$\epsilon H/\Omega$. In the problem of the charged particle
the adiabatic invariant is the series of the magnetic moment. 
Lewis's results came out from a direct application of the asymptotic theory of
Kruskal (1962) to the classical system described by $H_{L}$ with real
$\Omega$. Lewis found that Kruskal's theory could be applied in exact form. As a
consequence, an exact invariant, which is precisely the Ermakov-Lewis invariant,
has been found as a special case of Kruskal's adiabatic invariant. 
Although $\Omega$ was originally supposed to be real, the final results hold
for complex $\Omega$ as well. Moreover, the exact invariant is a constant of 
motion of the quantum system whose Hamiltonian is given by the quantum version
of eq.~(\ref{l1}).

\bigskip
\bigskip
\underline{The classical case.}

\bigskip
\noindent
Let us take a real $\Omega$. In order to correctly apply  
Kruskal's theory, it is necessary to write the equations of motion
as for an autonomous system of first order so that all solutions be periodic
in the independent variables for
$\epsilon\rightarrow 0$. This can be achieved by means 
of a new independent variable $s$ 
defined as $s=t/\epsilon$ and formally considering $t$ as a dependent variable. 
The resulting system of equations is
\begin{eqnarray}  \nonumber
dq/ds  &=& p~,\\  \nonumber
dp/ds &=& -\Omega ^2(t)q~,\\ \label{l2}
dt/ds &=& \epsilon\\   \nonumber
\end{eqnarray}
Since $t$ is now a dependent variable, this system is autonomous.
In the limit $\epsilon\rightarrow 0$, the solution of the last equation is 
$t={\rm constant}$, and therefore the other two equations
correspond to a harmonic oscillator of constant frequency.
Since $\Omega$ is real, the dependent  
variables are periodic in $s$ of period 
$2\pi /\Omega(t)$ in the limit $\epsilon\rightarrow 0$, and the system of 
equations has the form required by Kruskal's asymptotic theory. 
A central characteristic of the latter theory is a transformation from the 
variables $(q,p,t)$ to the so-called ``nice variables" 
$(z_1, z_2,\varphi)$. The latter are chosen in such a way that
a two-parameter family of closed curves in the space
$(q,p,t)$ can be defined by the conditions $z_1=$ 
constant and $z_2=$ constant. These closed curves are called rings. The 
variable $\varphi$ is a variable angle which is defined in such a way as to 
change by $2\pi$ if any of the rings is covered once. 
The rings have the important feature that all the family can be mapped 
to itself if on each ring $s$ is changed according to eqs.~(\ref{l2}). In the 
general theory, the transformation from the 
variables $(q,p,t)$ to the  
variables $(z_1,z_2,\varphi)$ is defined as an asymptotic series in 
positive powers of $\epsilon$, and a general prescription is given to determine
the transformation order by order.
As a matter of fact, Lewis has shown one possible explicit form for this 
transformation in terms of the variables $q$, $p$ and Pinney's function 
$\rho(t)$. Moreover, the inverse transformation can also be obtained in 
explicit form.

For the parametric oscillator problem, the rings are to be found in the 
$t$= constant planes. It is this property that allows the usage of the rings
for defining the exact invariant $I$ as the action integral
\begin{equation}  \label{l3}
I=\oint _{{\rm ring}}pdq~.
\end{equation}

By doing explicitly the integral of $I$ as an integral from  
$0$ to $2\pi$ over the variable $\varphi$ (see the Appendix), one gets
\begin{equation} \label{l4}
I=\frac{1}{2}[(q^2/\rho ^2)+(\rho \dot{x}-\epsilon\dot{\rho}q)^2]~,
\end{equation}
where $\rho$ satisfies Pinney's equation
\begin{equation} \label{l5}
\epsilon ^2\ddot{\rho}+\Omega ^2(t)\rho-1/\rho ^{3}=0~,
\end{equation}
and the point denotes differentiation with respect to $t$. 
The function $\rho$ can be taken as any particular solution of eq.~(\ref{l5}).
Although $\Omega$ was supposed to be real, $I$ is an  
invariant even for complex $\Omega$. It is easy to check that $dI/dt=0$ for the 
general case of complex $\Omega$ by performing the derivation of eq.~(\ref{l4}),
using
eqs.~(\ref{l2}) to eliminate $dq/dt$ and $dp/dt$, and eq.~(\ref{l5}) to
eliminate $\ddot{\rho}$.

It might appear that the problem of solving
the system of linear equations given by eqs.~(\ref{l2}) has been merely
replaced by the problem of solving the nonlinear eq.~(\ref{l5}). This is 
however not so.
First, any particular solution of eq.~(\ref{l5}) can be used in the formula of
$I$ with all the initial conditions for the eqs.~(\ref{l2}). 
For the numerical work it is sufficient to find a particular solution for 
$\rho$. 
Second, the exact invariant has a simple and explicit dependence on the 
dynamical variables $p$ and $q$. Third, taking into account the fact that 
$\epsilon ^2$ is a factor for $\ddot{\rho}$ in  
eq.~(\ref{l5}), one can obtain directly a particular solution for
$\rho$ as a series of positive powers of $\epsilon ^2$. 
If $\Omega$ is real and the leading term of the series
is taken as  $\Omega ^{-1/2}$, 
then the corresponding series solution is just the adiabatic invariant expressed
as a series.
It is interesting to speculate if in practice it is more useful to
calculate $I$ by means of the solution written as a truncated series of 
$\rho$ or by the corresponding expression in series for 
$I$ truncated at the same power of $\epsilon$. 
Forth, one can also solve eq.~(\ref{l5}) to get
$\rho$ as a power series in $1/\epsilon ^2$ in terms of integrales.
Finally, with the result of eqs.~(\ref{l4}) and
(\ref{l5}), it is possible to get a better understanding of the nature of 
Kruskal's adiabatic invariant. Some progress in this regard can be found in the
following general discussion on $I$ and $\rho$.

By adding a constant factor, the invariant $I$ of eq.~(\ref{l4}) is the most 
general quadratic invariant of the system whose Hamiltonian given by 
eq.~(\ref{l1}) is also a homogeneous quadratic form in $p$ and $q$. This can be
seen by writing the invariant in terms of two linear independent solutions,
$f(t)$ and $g(t)$ of the parametric equation. 
If we write the generalized form of $I$ 
\begin{equation} \label{l4b}
I=\delta ^2[\rho ^{-2}q^2+(\rho p-\epsilon \dot{\rho}q)^2]~,
\end{equation}
where $\delta$ is an arbitrary constant, and compare this form
with that in terms of $f(t)$ y $g(t)$, then we can infer that the two 
invariants are identical if $\rho$ is given by
\begin{equation} \label{l6}
\rho = \gamma _{1}(\epsilon \alpha)^{-1}\Big[\frac{A^2}{\delta ^2}g^2
+\frac{B^2}{\delta ^2}f^2+2\gamma _{2}\Big[\frac{A^2B^2}{\delta ^4}-
(\epsilon w)^2
\Big]^{1/2}fg\Big]^{1/2}~,
\end{equation}
where $A$ and $B$ are arbitrary constants, while the constants $\alpha$, 
$\gamma _{1}$ and $\gamma _2$ are defined by
\begin{equation} \label{l7}
w = fg^{'}-gf^{'}~,\quad 
\gamma _1 = \pm 1~,\quad 
\gamma _2 = \pm 1~.  
\end{equation}
Since there are two arbitrary constants, this formula for
$\rho$ gives the general solution of  
eq.~(\ref{l5}) expressed in terms of $f$ and $g$.
Using this formula we can build $\rho$ explicitly for any $\Omega$ 
for which the eqs.~(\ref{l2}) can be solved in an exact manner.
By constructing $\rho$ in this way for special cases,
we can infer that the expansion of $\rho$ in a series of positive powers of
$\epsilon ^2$ is at least sometimes convergent. For example, if
$\Omega=bt^{-2n/(2n+1)}$, where $b$ is a constant and $n$ is any integer,
the series expansion is a polynomial in $\epsilon ^2$, and consequently
it is convergent with an infinite radius of convergence.

Once we have the explicit form of Kruskal's invariant, it is possible to find
a canonical transformation for which the new momentum is the invariant itself. 
If we denote the new coordinate by
$q_1$, the conjugated momentum by $p_1$, and the generating function by
$F$, then the results can be written as
$$
q_1=-{\rm tan}^{-1}[\rho ^2p/q-\epsilon \rho \dot{\rho}]~,
$$
$$
p_1=\frac{1}{2}[\rho ^{-2}q^2+(\rho p-\epsilon \dot{\rho}q)^2]~,
$$
$$
F=\frac{1}{2}\epsilon\rho^{-1}\dot{\rho}q^2\pm\rho ^{-1}q(2p_1-\rho^{-2}q^2)^2
\pm p_1\sin ^{-1}[\rho ^{-1}q/(2p_1)^{1/2}]+(n+\frac{1}{2})\pi p_1
$$
$$
\left(-\frac{\pi}{2}\leq \sin ^{-1}[\rho ^{-1}q/(2p_{1})^{1/2}]
\leq \frac{\pi}{2}~,
n={\rm integer}\right)~,
$$
$$
p=\frac{\partial F}{\partial q}~, \qquad q_1=\frac{\partial F}{\partial p_1}~,
$$
\begin{equation} \label{l8}
H_{\rm new}=
H+\frac{\partial F}{\partial t}=\frac{1}{\epsilon}
\rho ^{-2}p_1~.
\end{equation}
In the expression for $F$ the upper or lower signs 
are taken according to 
$p-\epsilon\rho ^{-1}\dot{\rho}q$ is greater or less than $0$.
One can see that 
$q_1$ is a cyclic variable in the new Hamiltonian, as it should be if
$p_1=I$ can be an exact invariant.

Moreover, Lewis noticed that the second order differential equation for
$q$, namely $\epsilon ^2d^2q/dt^2+\Omega ^2(t)q=0$, is of the same form as the 
1D Schr\"odinger equation, 
if $t$ is considered as the spatial coordinate and $q$ is taken as the 
wave function. 
For bound states, $\Omega$ is imaginary whereas for the continuous spectrum 
$\Omega$ is real. Thus, the $I$ invariant is a relationship between the wave 
function and its first derivative \cite{F96}. 

\bigskip
\bigskip
\underline{The quantum case.}

\bigskip
\noindent
Let us consider the quantum system with the same 
Hamiltonian $H_{L}$,
where $\hat{q}$ and $\hat{p}$ should fulfill now the commutation relations
\begin{equation} \label{l9}
[\hat{q},\hat{p}]=i\hbar~.
\end{equation}
We shall take $\rho$ as real, which is possible if $\Omega ^2$ is real.
Using the commutation relationships and the equation for $\rho$ it is easy to 
show that $\hat{I}$ is a quantum constant of motion, i.e., it can be an 
observble since it satisfies 
\begin{equation} \label{l10}
\frac{d\hat{I}}{dt}=\frac{\partial \hat{I}}{\partial t}
+\frac{1}{i\hbar}[\hat{I},\hat{H}]=0~.
\end{equation}
It follows that $I$ has eigenfunctions whose eigenvalues
are time-dependent. The eigenfunctions and eigenvalues of $\hat{I}$ 
can be found by a method which is similar to that used by  Dirac 
to find the eigenfunctions and eigenvalues of the harmonic oscillator
Hamiltonian.
First, we introduce the raising and lowering operators, 
$\hat{a}^{\dagger}$ and $\hat{a}$, defined by
$$
\hat{a}^{\dagger}=(1/\sqrt{2})[\rho ^{-1}\hat{q}-i(\rho\hat{p}-
\epsilon \dot{\rho}\hat{q})]~,$$
\begin{equation} \label{l11}
\hat{a}=(1/\sqrt{2})[\rho ^{-1}\hat{q}+i(\rho\hat{p}-\epsilon\dot{\rho}\hat{q})]~.
\end{equation}
These operators fulfill the relationships
$$
[\hat{a},\hat{a}^{\dagger}]=\hbar~,
$$
\begin{equation}  \label{l12}
\hat{a}\hat{a}^{\dagger}=\hat{I}+\frac{1}{2}\hbar~.
\end{equation}
The operator $\hat{a}$ acting on an eigenfunction of $\hat{I}$ gives rise to an 
eigenfunction of $\hat{I}$ whose eigenvalue is less by $\hbar$ 
with respect to the initial eigenvalue. 
Similarly, $\hat{a}^{\dagger}$ acting on an eigenfunction 
of $\hat{I}$ raises the eigenvalue by $\hbar$. Once these properties
are settled, the normalization of the eigenfunctions of 
$\hat{I}$ can be used to prove that the eigenvalues of $\hat{I}$ are 
$(n+\frac{1}{2})\hbar$, where $n$ is $0$ or a positive integer. If 
$|n\rangle$ denotes the normalized eigenfunction of $\hat{I}$ whose 
eigenvalue is $(n+\frac{1}{2})\hbar$, we can express the relationhip between
$|n+1\rangle$ and $|n\rangle$ as follows
\begin{equation} \label{l13}
|n+1\rangle =[(n+1)\hbar]^{-1/2}\hat{a}^{\dagger}|n\rangle~.
\end{equation}
The condition that determines the eigenstate whose eigenvalue is
$\frac{1}{2}\hbar$ is given by
\begin{equation} \label{l14}
\hat{a}|0\rangle=0~.
\end{equation}
Using these results one can calculate the expectation value of the 
Hamiltonian in an eigenstate $|n\rangle$. The result is
\begin{equation} \label{l15}
\langle n|\hat{H}|n\rangle=(1/2\epsilon)(\rho ^{-2}+\Omega ^2 \rho ^2 +
\epsilon ^2 \dot{\rho}^2)(n+\frac{1}{2})\hbar~.
\end{equation}
It is interesting to note that the expectation values of 
$\hat{H}$ are equally spaced at each moment and that the lowest value is 
obtained for $n=0$, i.e., we have an exact counterpart of
the harmonic oscillator.
As a matter of fact, we can obtain the harmonic oscillator results
if $\Omega$ is taken real and constant with $\rho=\Omega ^{-1/2}$, which gives 
$I=\epsilon H/\Omega$.

\setcounter{equation}{0} 
\section*{6. The interpretation of Eliezer and Gray.}

The harmonic linear motion corresponding to the 1D parametric oscillator 
equation can be seen as the projection of a 2D motion of a particle driven
by the same law of force.
Thus, the 2D auxiliary motion is described by the equation 
\begin{equation} \label{eg1}
\frac{d^2\vec {r}}{dt^2}+\Omega^2 \left( t\right)\vec {r}=0
\end{equation}
where $\vec r$ is expressed in Cartesian coordinates $(x,y)$. Using polar 
coordinates $(\rho ,\theta)$ where $\rho = \vert r\vert $, 
$x=\rho \cos \theta $, $y=\rho\sin \theta$. The radial and transversal motions  
are described now by the equations  
\begin{equation} \label{eg2}
\ddot {\rho}-\rho {\dot\theta}^2 +\Omega^2 \rho =0
\end{equation}
\begin{equation}    \label{eg3}
\frac{1}{\rho}\frac{d}{dt}\left(\rho^2 \dot\theta\right)=0~.
\end{equation}
Integrating eq.~(\ref{eg3})
\begin{equation} \label{eg4}
\rho^2 \dot\theta =h
\end{equation}
where $h$ is the angular momentum, which is constant. Substituting in
eq.~(\ref{eg2}) one gets a Pinney equation of the form:
\begin{equation} \label{eg5}
\ddot {\rho}+\Omega^2 \rho =\frac{h^2}{\rho^3}
\end{equation}
The invariant $I$ corresponding to the eq.~(\ref{eg5}) is:
\begin{equation} \label{eg6}
I=\frac{1}{2}\left[ \frac{{h^2}{x^2}}{\rho^2} +\left(p\rho -x\dot \rho \right)^2\right]
\end{equation}
and with the substitutions $x=\rho \cos \theta $ and $p=\dot x$ one gets:
\begin{equation}
I=\frac{1}{2}\left[ {h^2}\cos^2\theta +{h^2}\sin^2\theta\right] =\frac{1}{2} h^2
\end{equation}
Thus, the constancy of $I$ is equivalent to the constancy of the auxiliary 
angular momentum.

In the elementary classical mechanics, the study of the simple 1D harmonic 
oscillator is often made as the projection of the uniform circular motion
on one of its diameters. The auxiliary motion introduced by Eliezer and Gray
is just a generalization of this elementary procedure to more general laws of 
force.

The connection between the solutions of the parametric oscillator linear 
equation and Pinney's solution is given by the following theorem. 

\bigskip

{\bf Theorem 1EG.} 
{\footnotesize 
{\em If $y_{\rm 1}$ and $y_{\rm 2}$ are linear independent solutions
of the equation
\begin{equation}
\frac{d^2 y}{dx^2} + Q\left( x\right) y=0
\end{equation}
and $W$ is the Wronskian $y_{\rm 1}y'_{\rm 2}-y_{\rm 2}y'_{\rm 1}$ (which,
according to Abel's theorem is constant), then the general solution of
\begin{equation}
\frac{d^2 y}{dx^2} + Q\left( x\right) y=\frac{\lambda}{y^3}
\end{equation}
where $\lambda$ is a constant, can be written as follows
\begin{equation} \label{pin}
y_{P}=\left( A{y_{\rm 1}}^2+B{y_{\rm 2}}^2+2Cy_{\rm 1}y_{\rm 2}\right)^{1/2}
\end{equation}
where $A$, $B$ and $C$ are constants such that
\begin{equation}
AB-C^2=\frac{\lambda}{W^2}
\end{equation}
}}

\bigskip

However, it is necessary that these constants be consistent with the initial
conditions of the motion.
If  
$x_{\rm 1}\left( t\right)$ and $x_{\rm 2}\left( t\right)$ are linear independent 
parametric solutions of initial conditions 
$x_{\rm 1}\left( 0\right) =1$, $\dot x_{\rm 1}\left( 0\right) =0$, 
$x_{\rm 2}\left( 0\right) =0$, $\dot x_{\rm 2}\left( 0\right) =1$, the
general parametric solution can be written as 
\begin{equation}\label{sgen}
x\left( t\right)=\alpha x_{\rm 1}\left( t\right) +\beta x_{\rm 2}\left( t\right)
\end{equation}
where $\alpha$ and $\beta$ are arbitrary constants
that are related to the initial conditions of the motion
by $x\left( 0\right)=\alpha$ and 
$\dot x\left( 0\right)=\beta$. The corresponding initial conditions for
$\rho$ and $\dot\rho$ are obtained from $x=\rho\cos\theta $, 
$\dot x=\dot\rho\cos\theta-\rho\dot\theta\sin\theta$, where 
$\theta \left( 0\right) =0$ gives $\rho \left( 0\right) =\alpha $ and 
$\dot\rho\left( 0\right)=0$. Using (\ref{pin}) we get
\begin{equation}
\rho\left( t\right) =\left[\left(\alpha x_{\rm 1}+\beta x_{\rm 2}\right)^2 +
\left(\frac{h^2}{\alpha^2} x_{\rm 2}^2\right)\right]^\frac{1}{2}
\end{equation}
as the solution of (\ref{eg5}) corresponding to the general parametric solution
(\ref{sgen}). 
Moreover, we have
\begin{equation}
\rho\cos\theta=\alpha x_{\rm 1}+\beta x_{\rm 2}
\end{equation}
\begin{equation}
\rho\sin\theta=\frac{hx_{\rm 2}}{\alpha}
\end{equation}

The previous considerations can be extended to systems whose equations 
of motion are of the form
\begin{equation} \label{PQ}
\frac{d^2 x}{dt^2}+P\left( t\right)\frac{d x}{dt}+Q\left( t\right) x=0~.
\end{equation}
The $I$ invariant is now 
\begin{equation}
I=\frac{h^2 x^2}{\rho^2}+\left(\dot\rho x-\rho p\right)^2 
\exp\left(2\int_{0}^{t} P\left( t\right) dt\right)
\end{equation}
where $\rho$ is any solution of 
\begin{equation} \label{PQp}
\frac{d^2 \rho}{dt^2}+P\left( t\right)\frac{d\rho}{dt}+Q\left( t\right)\rho=
\frac{h^2}{\rho^3}\exp\left(-2\int_{0}^{t} P\left( t\right) dt\right)
\end{equation}
The theorem that connects the solutions of (\ref{PQ}) with those of (\ref{PQp}) 
(with a change of notation) can be formulated in the following way. 

\bigskip

{\bf Theorem 2EG.} {\footnotesize {\em If 
$y_{1}\left( x\right)$ and $y_{2}\left( x\right)$ are two linear independent
solutions of
\begin{equation}
\frac{d^2 y}{dx^2}+P\left( x\right)\frac{d y}{dx}+Q\left( x\right) y=0
\end{equation}
the general solution of
\begin{equation}
\frac{d^2 y}{dx^2}+P\left( x\right)\frac{d y}{dx}+Q\left( x\right) y=
\frac{\lambda}{y^3}\exp\left(-2\int P\left( t\right) dt\right)
\end{equation}
can be written down as 
\begin{equation}
y=\left( A{y_{\rm 1}}^2+B{y_{\rm 2}}^2+2Cy_{\rm 1}y_{\rm 2}\right)^{1/2}
\end{equation}
where $A$ and $B$ are arbitrary constants, and
\begin{equation}
AB-C^2=\frac{\lambda}{W^2}\exp\left(-2\int P\left( t\right) dt\right)
\end{equation}
}}

\setcounter{equation}{0} 
\section*{ 7. The connection between the Ermakov invariant and N\"other's 
theorem.}

In 1978, Leach \cite{Le} found the Ermakov-Lewis invariant for the 
aforementioned parametric equation with first derivative
\begin{equation} \label{n1}
\ddot{x}+g(t)\dot{x}+\omega ^2(t)x=0~,
\end{equation}
by making use of a time-dependent canonical transformation leading to a constant
new Hamiltonian. That transformation belonged to a symplectic group and has been 
put forth without details.  
In the same 1978 year, Lutzky \cite{L78} proved that the invariant could be 
obtained starting from a direct application of Noether's theorem (1918).
This famous theorem makes a connection between the conserved quantities of a 
Lagrangian system with the group of symmetries that preserves the action
as an invariant functional. 
Moreover, Lutzky discussed the relationships between the solutions of the 
parametric equation of motion and Pinney's solution and commented on the great
potential of the method for solving the nonlinear equations. 

For the parametric equation without first derivative 
Lutzky used the following formulation of Noether's theorem.

\bigskip

{\bf Theorem NL.} {\footnotesize {\em Let $G$ be the one-parameter Lie group 
generated by 
$$
G=\xi(x,t)\frac{\partial}{\partial t}+n(x,t)\frac{\partial}{\partial x}
$$
such that the action functional $\int L(x, \dot{x}, t)dt$ is left
invariant under $G$. Then
\begin{equation} \label{n2}
\xi\frac{\partial L}{\partial t}+n\frac{\partial L}{\partial x}+
(\dot{n}-\dot{x}\dot{\xi})\frac{\partial L}{\partial \dot{x}}+\dot{\xi}L=
\dot{f}~.
\end{equation}
where $f=f(x,t)$, and
$$
\dot{\xi}=\frac{\partial \xi}{\partial t}+\dot{x}\frac{\partial \xi}
{\partial x}~, \qquad \qquad \dot{n}=
\frac{\partial n}{\partial t}+\dot{x}\frac{\partial n}
{\partial x}~, \qquad \qquad \dot{f}=
\frac{\partial f}{\partial t}+\dot{x}\frac{\partial f}
{\partial x}~.
$$
}}

\bigskip

Moreover, a constant of motion of the system is given by
\begin{equation} \label{n3}
\Phi=(\xi \dot{x}-n)\frac{\partial L}{\partial \dot{x}}-\xi L +f~.
\end{equation}
The 
Lagrangian $L=\frac{1}{2}(\dot{x}^2-\omega ^2 x^2)$ gives the equations of 
motion of the parametric oscillating type; substituting this
Lagrangian in (\ref{n2}) and equating to zero the coefficients of the 
corresponding 
powers of $\dot{x}$, on gets a set of equations for
$\xi$, $n$, $f$. Next, it is easy to prove that they imply that $\xi$ 
is a function only of $t$ and fulfills



\begin{equation} \label{n4}
\stackrel{...}{\xi}+4\xi \omega \dot{\omega}+4\omega ^2\dot{\xi}=0~.
\end{equation}
The following results are easy to get
$$
n(x,t)=\frac{1}{2}\dot{\xi}x+\psi(t)~,
$$
$$
f(x,t)=\frac{1}{4}\ddot{\xi}x^2+\dot{\psi}x+C~, \qquad \ddot{\psi}+\omega ^2
\psi=0~.
$$

Choosing $C=0$, $\psi=0$, and substituting these values in (\ref{n3}), 
one can find that
\begin{equation}  \label{n5}
\Phi=\frac{1}{2}(\xi \dot{x}^2+[\xi \omega ^2+\frac{1}{2}\ddot{\xi}]x^2-
\dot{\xi}x\dot{x})
\end{equation}
is a conserved quantity for the parametric undamped oscillatory motion
if $\xi$ satisfies (\ref{n4}). Notice that (\ref{n4}) 
has the first integral
\begin{equation} \label{n6}
\xi\ddot{\xi}-\frac{1}{2}\dot{\xi}^2 +2\xi ^2\omega ^2=C_1~.
\end{equation}
If we choose $\xi=\rho ^2$ in (\ref{n5}) and (\ref{n6}), with $C_1=1$, 
we get that $\Phi$ is the  Ermakov-Lewis invariant. If the formula for the 
latter is considered 
as a differential equation for $x$, then it is easy to solve in the  
variable $x/\rho$; the result can be written in the form 
\begin{equation} \label{n7}
x=\rho[A\cos \phi +B\sin \phi]~, \qquad \phi=\phi(t)~,
\end{equation}
where $\dot{\phi}=1/\rho ^2$ and $A$ and $B$ are arbitrary constants. 
Thus, the general parametric solution can be found if a particular solution of
Pinney's equation is known. 

Consider now the Ermakov-Lewis invariant as a 
conserved quantity for Pinney's equation; this is possible if  
$x$ fulfills the parametric equation of motion. This standpoint is interesting 
because it provides an example of how to use 
Noether's theorem to change a problem of solving nonlinear equations
into an equivalent problem of solving linear equations. 
Thus, if we take as our initial task to solve Pinney's equation, 
we can use Noether's theorem with
$$
L(\rho, \dot{\rho}, t)=\frac{1}{2}(\dot{\rho}^2-\omega ^2 \rho ^2 -
\frac{1}{\rho ^2})~,
$$
to prove that
\begin{equation} \label{n8}
\Phi =\frac{1}{2}\Big[\frac{x^2}{\rho ^2}+C_2\frac{\rho ^2}{x^2}+
(\rho \dot{x}-\dot{\rho}x)^2\Big]
\end{equation}
is a conserved quantity for Pinney's equation leading to
\begin{equation} \label{n9}
\ddot{x}+\omega ^2x=C_2/x^3~.
\end{equation}
The quantity $C_2$ is an arbitrary constant; choosing $C_2=0$, we reduce 
Pinney's solution to the parametric linear solution, while 
(\ref{n8}) turns into the Ermakov-Lewis invariant.

If we write the invariant for two different solutions of the linear parametric 
equation, $x_1$ and $x_2$,
while keeping the same $\rho$, and eliminate $\dot{\rho}$ in the resulting 
equations we get
\begin{equation} \label{n10}
\rho=\frac{1}{W}\sqrt{I_1x_2^2+I_2x_1^2+2x_1x_2[I_1I_2-W^2]^{1/2}}~,
\end{equation}
where $W=\dot{x}_1x_2-x_1\dot{x}_2$, and $I_1$ and $I_2$ are constants. 
Thus, a general solution of Pinney's equation can be obtained if
two solutions of the linear parametric equation can be found.
(Since the Wronskian $W$ is constant for two independent linear solutions,
we can find that $I_1=1$, $I_2=W^2$, and therefore (\ref{n10}) 
turns into $\rho=\sqrt{x_1^2+(1/W^2)x_2^2}$, which is the result given by
Pinney in 1950). Moreover, one can see from (\ref{n7}) that two independent
parametric solutions are $x_1=\tilde{\rho}\cos \phi$,
$x_2=\tilde{\rho}\sin \phi$, where $\tilde{\rho}$ is any solution of 
Pinney's equation. Then $W=1$, and (\ref{n10}) turns into
\begin{equation} \label{n11}
\rho=\tilde{\rho}\sqrt{I_1\sin ^2\phi+I_2\cos ^2\phi+[I_1I_2-1]^{1/2}
\sin 2\phi}~,\qquad \qquad \dot{\phi}=1/\tilde{\rho}^2~.
\end{equation}
This beautiful result obtained by Lutzky by means of Noether's theorem gives the
general solution of Pinney's equation in terms of an arbitrary particular 
solution of the same equation. Moreover, Lutzky suggested that this approach can
be used to solve certain nonlinear dynamical systems once a conserved quantity
containing an auxiliary function of a corresponding nonlinear differential
equation can be found. 
Even if the auxiliary equation is nonlinear, sometimes it is simpler to solve
than the original linear equation. In any case, one can establish useful
relationships between the solutions of the two types of equations.

In conclusion, we mention that Noether's method can be applied 
to the equation of parametric motion with first derivative 
(\ref{n1}); in this way one can reproduce the results of  
Eliezer and Gray of the previous chapter. 
The effective Lagrangian for (\ref{n1}) is given by
$L=\frac{1}{2}e^{F(t)}[\dot{x}^2-\omega ^2(t)x^2]$, where $dF/dt=g(t)$.

\setcounter{equation}{0} 
\section*{ 8. Possible generalizations of the Ermakov method.}

We have seen that there is a simple relationship between the solutions of the 
parametric oscillator
\begin{equation}  \label{rr1}
\ddot{x}+\omega ^2(t)x=0~,
\end{equation}
and the solution of nonlinear differential equations of the Pinney type
that differ from eq.~(\ref{rr1}) only in the nonlinear term. 
The equation of motion of a charged particle in some types of time-dependent
magnetic fields can be written in the above form. Many time-dependent 
oscillating systems are governed by the same eq.~(1). A conserved quantity for
eq.~(\ref{rr1}) is
\begin{equation}  \label{rr2}
I_{EL}=\frac{1}{2}[(x^2/\rho ^2)+(\rho \dot{x}-\dot{\rho}x)^2]~,
\end{equation}
where $x(t)$ satisfies eq.~(\ref{rr1}) and $\rho(t)$ satisfies the auxiliary
equation 
\begin{equation} \label{rr3}
\ddot{\rho}+\omega ^2(t)\rho=1/\rho ^{3}~.
\end{equation}

Using eq.~(\ref{rr1}) to eliminate $\omega ^{2}(t)$ in eq.~(\ref{rr3}) 
we find
\begin{equation} \label{rr4}
\ddot{\rho}+(\rho/x)\ddot{x}=1/\rho ^3~,
\end{equation}
or
\begin{equation} \label{rr5}
x\ddot{\rho}-\rho\ddot{x}=(d/dt)(x\dot{\rho}-\rho\dot{x})=x/\rho^{3}~.
\end{equation}
Now, multiplying this equation by $x\dot{\rho}-\rho\dot{x}$ we can write 
\begin{equation} \label{rr6}
(x\dot{\rho}-\rho\dot{x})(d/dt)(x\dot{\rho}-\rho\dot{x})=
(x\dot{\rho}-\rho\dot{x})x/\rho ^{3}~,
\end{equation}
or
\begin{equation} \label{rr7}
\frac{1}{2}(d/dt)(x\dot{\rho}-\rho\dot{x})^{2}=-\frac{1}{2}(d/dt)
(x/\rho ^{2})~,
\end{equation}
and therefore we have the invariant 
\begin{equation}  \label{rr8}
I_{EL}=\frac{1}{2}[(x^2/\rho ^2)+(\rho \dot{x}-\dot{\rho}x)^2]~,
\end{equation}
where $x$ is any solution of eq.~(\ref{rr1}) and $\rho$ is any solution
of eq.~(\ref{rr3}).

A simple generalization of this result has been proposed by Ray and Reid in 1979
\cite{RR}. Instead of (\ref{rr3}) they considered the following equation 
\begin{equation}  \label{rr9}
\ddot{\rho}+\omega ^{2}(t)\rho=(1/x\rho ^2)f(x/\rho)~,
\end{equation}
where $x$ is a solution of eq.~(\ref{rr1}) and $f(x/\rho)$ is an 
{\em arbitrary} function of $x/\rho$. If again we eliminate $\omega ^2$ and we
employ as a factor $x\dot{\rho}-\rho\dot{x}$ as a factor to obtain
\begin{equation}  \label{rr10}
\frac{1}{2}(d/dt)(x\dot{\rho}-\rho\dot{x})^2=-(d/dt)\phi(x/\rho)~,
\end{equation}
where
\begin{equation} \label{rr11}
\phi(x/\rho)=2\int ^{x/\rho}f(u)du~.
\end{equation}
From  eq.~(\ref{rr10}) we have the invariant
\begin{equation} \label{rr12}
I_{f}=\frac{1}{2}[\phi(x/\rho)+(\rho \dot{x}-\dot{\rho}x)^2]~,
\end{equation}
where $x$ is a solution of eq.~(\ref{rr1}) and $\rho$ is a solution of 
eq.~(\ref{rr9}). For $f=x/\rho$ we reobtain the invariant $I_{EL}$. The result
(\ref{rr12}) provides a connection between the solutions of the linear equation
(\ref{rr1}) with the solutions of an infinite number of nonlinear equations
(\ref{rr9}) by means of the invariant $I_{f}$.

As an additional generalization, one can consider the following two
equations 
\begin{equation} \label{rr13}
\ddot{x}+\omega ^2(t)x=(1/\rho x^2)g(\rho/x)~,
\end{equation}
\begin{equation} \label{rr14}
\ddot{\rho}+\omega ^2(t)\rho=(1/x\rho ^2)f(x/\rho)~,
\end{equation}
where $g$ and $f$ are {\em arbitrary functions} of their arguments. 
Applying the same procedure to these equations one can find the invariant
\begin{equation} \label{rr15}
I_{f,g}=\frac{1}{2}[\phi(x/\rho)+\theta(\rho/x)+(x\dot{\rho}-\rho\dot{x})^2]~,
\end{equation}
where
\begin{equation} \label{rr16}
\phi(x/\rho)=2\int ^{x/\rho}f(u)du~,
\end{equation}
\begin{equation} \label{rr17}
\theta(\rho/x)=2\int ^{\rho/x}g(u)du~.
\end{equation}
The expression
(\ref{rr15}) is an invariant whenever $x$ is a solution of  
eq.~(\ref{rr13}) and $\rho$ is a solution of eq.~(\ref{rr14}). One should
notice that the functions $f$ and $g$ are arbitrary, and therefore the invariant 
$I_{f,g}$ gives the connection between the solutions of many different 
differential equations. We can see that the  
Ermakov-Lewis invariant is merely a particular case of $I_{f,g}$ with 
$g=0$, $f=x/\rho$.

In the cases $g=0$, $f=0$; $g=0$, $f=x/\rho$; $g=\rho/x$, $f=0$; and $f=x/\rho$,
$g=\rho/x$ the equations (\ref{rr13}) and (\ref{rr14}) respectively are not
coupled. In general, if we have found a solution for $x$, then the 
invariant $I_{f,g}$ provides some information about the solution
$\rho$.

On the other hand, it is not known if the simple mechanical interpretation
of Eliezer and Gray is also available for different choices of $f$ and $g$. 
The simple proof of the existence of $I_{f,g}$ clarifies how such invariants 
can occur from pairs of differential equations. 

\setcounter{equation}{0} 
\section*{ 9. Geometrical angles and phases in the Ermakov problem.}

The quantum mechanical holonomic effect known as 
Berry's phase (BP) (1984) has been of much interest in the last fifteen years. 
In the simplest cases, it shows up when the time-dependent parameters of a 
system change adiabatically in time in the course of a closed trajectory in the 
parameter space. 
The wave function of the system gets, in addition to the common dynamical phase 
$\exp(-i\hbar \int_{0}^{T}E_{n}(t)dt)$, a geometrical phase factor given by
\begin{equation}  \label{dam1}
\gamma _{n}(c)=i\int _{0}^{T}dt \langle \Psi _{n}(X(t))|\frac{d}{dt}|
\Psi _{n}(X(t))\rangle~,
\end{equation}
because the parameters are slowly changing along the closed path $c$ 
of the spatial parameter $X(t)$ during the period $T$. 
$|\Psi _{n}(X(t))\rangle$ are the eigenfunctions
of the instantaneous Hamiltonian $H(X(t))$. BP has a classical analogue as an
angular shift accumulated by the system when its dynamical   
variables are expressed in angle-action variables. This angular shift is known in
the literature as Hannay's angle 
(Hannay 1985, Berry 1985). Various model systems have been employed to calculate
the BP and its classical analog. One of these systems is the generalized 
harmonic oscillator whoose $H$ is given by 
(Berry 1985, Hannay 1985)
\begin{equation}  \label{dam2}
H_{XYZ}(p,q,t)=\frac{1}{2}[X(t)q^2+2Y(t)qp+Z(t)p^2]~,
\end{equation}
where the slow time-varying parameters are $X(t)$, $Y(t)$ y $Z(t)$.

Since $H_{XYZ}$ can be transformed into the 
$H$ of a parametric oscillator, it follows that there should be a connection 
between the BP of the system 
with $H_{XYZ}$ and the Lewis phase for the parametric oscillator \cite{LR}. 
This problem has been first studied by Morales \cite{mor}.
Interestingly, the results appear to be exact although the system does not 
evolve adiabatically in time and goes to Berry's result in the adiabatic limit. 

Lewis and Riesenfeld \cite{LR} have shown that for 
a quantum nonstationary system which is characterized by a  
Hamiltonian $\hat{H}(t)$ and a Hermitian invariant $\hat{I}(t)$, the general 
solution of the Schroedinger equation 
\begin{equation} \label{dam3}
i\hbar \frac{\partial \Psi(q,t)}{\partial t}=\hat{H}(t)\Psi(q,t)~,
\end{equation}
is given by
\begin{equation} \label{dam4}
\Psi(q,t)=\sum _{n}C_{n}\exp (i\alpha _{n}(t))\Psi _{n}(q,t)~.
\end{equation}
$\Psi _{n}(q,t)$ are the eigenfunctions of the invariant
\begin{equation} \label{dam5}
\hat{I}\Psi _{n}(q,t)=\lambda _{n}\Psi _{n}(q,t)~,
\end{equation}
where the eigenvalues are time-dependent, the coefficients $C_{n}$ are 
constants and the phases $\alpha _{n}(t)$ are obtained from the equation
\begin{equation} \label{dam6}
\hbar d\alpha _{n}(t)/dt=\langle\Psi_{n}|i\hbar \partial/\partial t-\hat{H}(t)|
\Psi _{n}\rangle ~.
\end{equation}
Using this result, Lewis and Riesenfeld obtained solutions for a quantum harmonic
oscillator of parametric frequency characterized by the classical Hamiltonian 
\begin{equation} \label{dam7}
H(t)=\frac{1}{2}p^2+\frac{1}{2}\Omega ^{2}(t)q^2
\end{equation}
and the classical equation of motion
\begin{equation}  \label{dam8}
\ddot{q}+\Omega ^{2}(t)q^2=0~,
\end{equation}
where the points denote differentiation with respect to time. The matrix 
elements that are required to evaluate the BP are given by \cite{LR}
\begin{equation}  \label{dam9}
\langle \Psi _{n}|\partial/\partial t|\Psi _{n}\rangle =\frac{1}{2}i
(\rho \ddot{\rho}-\dot{\rho}^{2})(n+\frac{1}{2})~.
\end{equation}
\begin{equation}  \label{dam10}
\langle \Psi _{n}|\hat{H}(t)|\Psi _{n}\rangle =\frac{1}{2}
(\dot{\rho}^{2}+\Omega ^{2}(t)\rho ^{2}+1/\rho ^{2})(n+\frac{1}{2})~,
\end{equation}
where $\rho (t)$ is a real number, satisfying the equation
\begin{equation} \label{dam11}
\ddot{\rho}+\Omega ^{2}(t)\rho =1/\rho ^3~.
\end{equation}
Substituting (\ref{dam9}) and (\ref{dam10}) in (\ref{dam6}) and 
integrating one gets
\begin{equation} \label{dam12}
\alpha _{n}(t)=-(n+\frac{1}{2})\int _{0}^{t}dt^{'}/\rho ^{2}(t^{'})~.
\end{equation}
One can show this either by using (\ref{dam9}) or (\ref{dam12}) and one can get  
the BP and Hannay's angle for the system of
 Hamiltonian $H_{XYZ}$. For this system, the frequency that can be obtained  
from the Hamiltonian expressed in the action-angle variables, is given by
\begin{equation} \label{dam13}
\omega =\partial H(I,X(t),Y(t),Z(t))/\partial I=(XZ-Y^2)^{1/2}~.
\end{equation}
From (\ref{dam2}) one can get the equations of motion for $q$ and $p$ and by 
eliminating $p$ one can get the Newtonian equation of motion for $q$ as follows
\begin{equation} \label{dam14}
\ddot{q}-(\dot{Z}/Z)\dot{q}+[XZ-Y^2+(\dot{Z}Y-\dot{Y}Z)/Z]q=0~.
\end{equation}
The term in $\dot{q}$ can be eliminated by introducing a new coordinate
$Q(t)$ given by (Berry 1985)
\begin{equation} \label{dam15}
q(t)=[Z(t)]^{1/2}Q(t)~.
\end{equation}
Substituting (\ref{dam15}) in (\ref{dam14}) one gets
\begin{equation} \label{dam16}
\ddot{Q}+[XZ-Y^2+(\dot{Z}Y-\dot{Y}Z)/Z+[1/2(\ddot{Z}/Z-\dot{Z}^2/Z^2)
-1/4(\dot{Z}/Z)^2]]Q=0~,
\end{equation}
which corresponds to the equation of motion of an oscillator with parametrically
forced frequency. 
Berry found Hannay's angle 
$\Delta \theta$ by the WKB method in quantum mechanics, 
but it can also be obtained 
by means of (\ref{dam9}) or (\ref{dam12}).

Comparing (\ref{dam8}) with (\ref{dam16}) we see that we can define 
$\Omega ^2(t)$ as
\begin{equation} \label{dam17}
\Omega ^{2}(t)=XZ-Y^2+(\dot{Z}Y-\dot{Y}Z)/Z+[1/4(\ddot{Z}/Z-\dot{Z}^2/Z^2)-
1/4(\dot{Z}/Z)^2]~.
\end{equation}
With this connection and employing (\ref{dam1}) and (\ref{dam9}) we get
\begin{equation}  \label{dam18}
\gamma _{n}(C)=-\frac{1}{2}(n+\frac{1}{2})\int _{0}^{T}(\rho \ddot{\rho}-
\dot{\rho}^2)dt~,
\end{equation}
where $\rho(t)$ is the solution of (\ref{dam11}) with $\Omega ^2(t)$ given by 
(\ref{dam17}). It is important to notice that (\ref{dam18}) is 
{\em exact} even when the system does not evolve slowly in time.

To compare with Berry's result one should take the adiabatic limit.
For that we define an adiabaticity parameter $\epsilon$ and a 
 `slow time' variable $\tau$
\begin{equation}  \label{dam19}
\xi\equiv \xi(\tau)\qquad \tau=\epsilon t~,
\end{equation}
in terms of which $\Omega ^2(\tau)$ turns into
\begin{equation} \label{dam20}
\Omega ^2(\tau)=\epsilon ^{-2}[XZ-Y^2+\epsilon(Z^{'}Y-Y^{'}Z)/Z+\epsilon ^{2}
[1/2(Z^{'}/Z)^{'}-1/4(Z^{'}/Z)^{2}]]~,
\end{equation}
where the primes indicate differentiation with respect to $\tau$. It has been 
shown by Lewis that in the adiabatic limit eq.~(\ref{dam20}) can be solved as a
power series in $\epsilon$ with the leading term given by
\begin{equation} \label{dam21}
\rho _{0}=\Omega ^{-1/2}(\tau)~.
\end{equation}
If we plug this expression for $\rho$ and its time derivatives in  
(\ref{dam18}) we could obtain the BP in the adiabatic limit.
However, it is easy to calculate the Lewis phase first and then resting it from
the dynamical phase
$-\hbar ^{-1}\langle \Psi _{n}|H(t)|\Psi _{n}\rangle$.
Substituting (\ref{dam20}) and (\ref{dam21}) in (\ref{dam12}) we get
\begin{equation} \label{dam22}
\alpha _{n}(\tau)=-(n+\frac{1}{2})\left(\frac{1}{\epsilon}\int _{0}^{\tau}
(XZ-Y^2)^{1/2}d\tau ^{'}+\frac{1}{2}\int _{0}^{\tau}\frac{(Z^{'}Y-Y^{'}Z)}
{Z(XZ-Y^2)^{1/2}}d\tau ^{'}+O(\epsilon)\right)~.
\end{equation}
The first term in the right hand side is the dynamical phase and the 
second and hiher order terms are associated with Berry's phase. Therefore we can
write the BP as follows
\begin{equation} \label{dam23}
\gamma _{n}(C)=-\frac{1}{2}(n+\frac{1}{2})\int _{0}^{T}
\frac{\dot{Z}Y-\dot{Y}Z}{Z(XZ-Y^2)^{1/2}}dt~.
\end{equation}
Hannay's angle is obtained by using the correspondence principle in the form 
(Berry 1985, Hannay 1985)
\begin{equation} \label{dam24}
\Delta \theta =-\partial \gamma _{n}/\partial n~,
\end{equation}
as
\begin{equation} \label{dam25}
\Delta \theta =\frac{1}{2}\int _{0}^{T}
\frac{(\dot{Z}Y-\dot{Y}Z)}{Z(XZ-Y^2)^{1/2}}dt~,
\end{equation}
which is the same result as that obtained by Berry (1985).

In this way, it has been proved that if a time-dependent quadratic $H$ 
can be transformed to the parametric form given by (\ref{dam7}), 
then the Lewis phase can be used to calculate the BP  and the Hannay's angle.
Although we presented the particular case discussed by Morales, it is known that
Lewis' approach for time-dependent systems is general.
As a matter of fact, one can find more general cases in the literature.

\setcounter{equation}{0} 
\section*{ 10. Application to minisuperspace Hamiltonian cosmology.}

The formalism of Ermakov invariants can be a useful alternative to study 
the evolutionary and chaoticity problems of ``quantum" canonical universes 
since these invariants are closely related to the Hamiltonian formulation. 
Moreover, as we have seen in the previous chapter, 
Ermakov's method is intimately related to geometrical angles and phases 
\cite{book}. Therefore, it seems natural to speak of  
Hannay's angle as well as of various types of topological phases at the 
cosmological level.

The  Hamiltonian formulation of the general relativity
has been developed in the classical works of Dirac \cite{D} and 
Arnowitt, Deser and Misner (ADM) \cite{ADM}. When it was applied to the Bianchi
homogeneous cosmological models it led to the so-called  
Hamiltonian cosmology \cite{lR}. Its quantum counterpart, 
the canonical quantum cosmology \cite{lE}, 
is based on the canonical quantization methods and/or path integral procedures.
These cosmologies are often used in heuristic studies of the very early 
universe, close to the Planck scale epoch $t_{P}\approx 10^{-43}$ s.

The most general models for homogeneous cosmologies are the Bianchi ones. 
In particular, those of class A of diagonal metric
are at the same time the simplest from the point of view of quantizing them.
 
Briefly, we can say that in the 
ADM formalism the metric of these models is of the form 
\begin{equation} \label{c1}
\rm{ ds^2= -dt^2 + e^{2\alpha(t)}\, (e^{2\beta(t)})_{ij}\, \omega^i \,
\omega^j,}
\end{equation}
where $\alpha(t)$ is a scalar function and $\rm \beta_{ij}(t)$ is a diagonal 
matrix of dimension 3, 
$\rm \beta_{ij}= diag(x+ \sqrt{3} y,x- \sqrt{3} y, -2x)$,
$\rm \omega^i$ are 1-forms characterizing each of the Bianchi models and
fulfilling the algebra 
$\rm d\omega^i= {1\over 2} C^i_{jk} \omega^j \wedge \omega^k$, where 
$\rm C^i_{jk}$ are structure constants.

The ADM action has the form
\begin{equation} \label{c2}
\rm {I_{ADM}=\int (P_x dx+ P_y dy + P_{\alpha} d\alpha -  
N {\cal H}_\perp) dt,}
 \end{equation}
where the Ps are the canonical moments, N is the lapse function
and
\begin{equation} \label{c3}
{\cal H}_\perp= \rm e^{-3\alpha}\left(-P^2_\alpha+ P^2_x +P^2_y +
e^{4\alpha}V(x,y) \right)~.
\end{equation}
$\rm e^{4\alpha}\, V(x,y)= U(q^{\mu})$ is the potential of the cosmological
model under consideration. The  
Wheeler-DeWitt (WDW) equation can be obtained by canonical quantization, i.e.,
substituting P$_{\rm{q^{\mu}}}$ by $\rm{\hat{P}_{q^{\mu}}= 
-i \partial_{q^{\mu}}}$ in eq.~(3), where 
$\rm{ q^{\mu}=(\alpha, x,  y)}$. The factor ordering of $\rm{ e^{-3\alpha}}$ 
with the operator $\rm {\hat P_{\alpha}}$ is not unique. Hartle and Hawking 
\cite{hh} suggested an almost general ordering of the following type 
\begin{equation} \label{c4}
\rm{ - e^{-(3- Q)\alpha}\, \partial _{\alpha} e^{-Q\alpha} \partial _{\alpha}
=
- e^{-3\alpha}\, \partial^2_{\alpha} + Q\, e^{-3\alpha} \partial _{\alpha},}
\end{equation}
where $Q$ is any real constant. If $Q=0$ the WDW equation is
\begin{equation} \label{c5}
\rm {\Box \, \Psi - U(q^{\mu}) \, \Psi =0,}
\end{equation}
Using the ansatz
$\rm \Psi(q^{\mu}) = A e^{\pm \Phi}$
one gets
\begin{equation}  \label{c6}
  \pm A {\Box \, \Phi}  
+ A [ (\nabla \Phi)^2 - U] = 0,
\end{equation}
where $\rm \Box = G^{\mu \nu}{\partial ^2\over \partial q^{\mu} \partial
 q^{\nu}}$,
$\rm (\nabla) ^2= -({\partial \over \partial \alpha})^2
+({\partial \over \partial x})^2 +
({\partial \over \partial y})^2$, y
$\rm G^{\mu \nu}= diag(-1,1,1)$. 

Employing the change of variable
$\rm (\alpha, x, y)\rightarrow (\beta _1, \beta _2, \beta _3)$, where
\begin{equation} \label{c7}
\beta _1 = \alpha + x + \sqrt 3 y,\qquad
\beta _2 = \alpha + x - \sqrt 3 y,\qquad
\beta _3 = \alpha - 2x,
\end{equation}
the 1D character of some of the Bianchi models can be studied in a more direct 
way.  

\bigskip
\bigskip

\underline{ Empty FRW (EFRW) universes for $Q=0$.}

\bigskip

\noindent
We now apply the Ermakov method to the simplest cosmological oscillators
which are the empty quantum universes of   
Friedmann-Robertson-Walker (EFRW) type.
The results included in this chapter have been published recently \cite{rer}.
When the Hartle-Hawking parameter is equal to zero ($Q=0$), the 
WDW equation for the EFRW universe is 
\begin{equation} \label{qc1}
\frac{d^2\Psi}{d\Omega^2} -\kappa e^{-4\Omega}\Psi
(\Omega) =0~,
\end{equation}
where $\Omega$ is the Misner time which is related to the volume of the universe 
$V$ at a given cosmological epoch as $\Omega=-\ln (V^{1/3})$ 
\cite{Mi}, $\kappa$ is the curvature parameter of the universe (1,0,-1 for 
closed, plane, open universes, respectively) and $\Psi$ is the wave function of 
the universe. The general solution is obtained as a linear superposition of 
modified Bessel functions of zero order
in the case for which $\kappa =1$, 
$\Psi(\Omega)=C_1I_{0}(\frac{1}{2}
e^{-2\Omega})+C_2K_{0}(\frac{1}{2}e^{-2\Omega})$. If $\kappa =-1$ the solution
will be a superposition of ordinary Bessel functions of zero order 
$\Psi(\Omega)=C_1J_{0}(\frac{1}{2}
e^{-2\Omega})+C_2Y_{0}(\frac{1}{2}e^{-2\Omega})$. $C_1$ and $C_2$ are arbitrary
superposition constants that we shall take for simplicity reasons equal 
$C_1=C_2=1$.

Eq.~(\ref{qc1}) can be transformed into the canonical equations of 
motion for a classical point particle of mass $M=1$, generalized coordinate
$q=\Psi$ and  moment 
$p=\Psi ^{'}$, and by considering Misner's time as a Hamiltonian time for which we 
shall keep the same notation. Thus, we can write
\begin{eqnarray}
\frac{dq}{d\Omega}&=&p~\\     
\frac{dp}{d\Omega}&=&\kappa e^{-4\Omega}q~.  
\end{eqnarray}
These equations describe the canonical motion of an inverted oscillator
($\kappa =1$) and of a normal one ($\kappa =-1$), respectively \cite{invert},
of Hamiltonian
\begin{equation} \label{4}
H(\Omega)=\frac{p^2}{2}-\kappa e^{-4\Omega}\frac{q^2}{2}~.
\end{equation}
For the EFRW Hamiltonian the phase space functions 
$T_1=\frac{p^2}{2}$, $T_2=pq$, y $T_3=\frac{q^2}{2}$ form a dynamical Lie 
algebra, i.e., 
\begin{equation}\label{4b}
H=\sum _{n}h_{n}(\Omega)T_{n}(p,q)~,
\end{equation}
which is closed with respect to the Poisson brackets 
$\{T_1,T_2\}=-2T_1$, $\{T_2,T_3\}=-2T_3$, $\{T_1,T_3\}=-T_2$. The Hamiltonian 
EFRW Hamiltonian can be written now as
\begin{equation} \label{H}
H=T_1-\kappa e^{4\Omega}T_3~.
\end{equation}
The Ermakov invariant $I$ is a function in the dynamical algebra 
\begin{equation} \label{5}
I=\sum _{r}\epsilon _{r}(\Omega)T_{r}~,
\end{equation}
and through the time invariance condition
\begin{equation} \label{6}
\frac{\partial I}{\partial \Omega}=-\{I,H\}~,
\end{equation}
one is led to the following equations for the unknown functions
$\epsilon _{r}(\Omega)$
\begin{equation} \label{7}
\dot{\epsilon} _{r}+\sum _{n}\Bigg[\sum _{m}C_{nm}^{r}h_{m}
(\Omega)\Bigg]\epsilon _{n}=0~,
\end{equation}
where $C_{nm}^{r}$ are the structure constants of the Lie algebra given above. 
Thus, we obtain
\begin{eqnarray} \nonumber
\dot{\epsilon} _1&=&-2\epsilon _2 \\
\dot{\epsilon} _2&=&-\kappa e^{-4\Omega}\epsilon _1-\epsilon _3\\
\dot{\epsilon} _3&=&-2\kappa e^{-4\Omega}\epsilon _2~.    \nonumber
\end{eqnarray}
The solution of this system of equations can be easily obtained by choosing
$\epsilon _1=\rho ^2$, which gives $\epsilon _2=-\rho \dot{\rho}$ y
$\epsilon _3=\dot{\rho} ^2 +\frac{1}{\rho ^2}$, where $\rho$ is the solution 
of Pinney's equation 
\begin{equation} \label{9}
\ddot{\rho}
-\kappa e^{-4\Omega}\rho=\frac
{1}{\rho ^3}~.
\end{equation}
In terms of $\rho (\Omega)$ and using (6), the 
Ermakov invariant can be written as follows 
\begin{equation} \label{10}
I=I_{{\rm kin}}+I_{{\rm pot}}=
\frac{(\rho p-\dot{\rho}q)^2}{2}+\frac{q^2}{2\rho ^2}=\frac{\rho ^4}{2}
\Big[\frac{d}{d\Omega}\left(\frac{\Psi}{\rho}\right)\Big]^2+\frac{1}{2}
\left(\frac{\Psi}{\rho}\right)^2.
\end{equation}
We have followed the calculations of Pinney and of Eliezer and Gray
for $\rho(\Omega)$ in terms of linear combinations 
of the aforementioned Bessel functions that satisfy
the initial conditions as given by these authors.
We have worked with the values $A=1$, $B=-1/W^2$ y $C=0$ for 
Pinney's constants, where $W$ is the Wronskian of the Bessel functions. 
We have also chosen an auxiliary angular moment of unit value ($h=1$). 
Since $I=h^2/2$,we have to obtain a constant value of one-half for the 
Ermakov invariant. We have checked this by plotting $I(\Omega)$ for 
$\kappa=\pm1$ in fig.~1.



Now we pass to the calculation of the angular variables. We first calculate
the time-dependent generating function that allows us to go to the new canonical 
variables for which $I$ is chosen as the new ``moment'' \cite{L}
\begin{equation} \label{11}
S(q,I,\vec{\epsilon}(\Omega))=\int ^{q}dq^{'}p(q^{'},I,\vec{\epsilon}
(\Omega))~,
\end{equation}
leading to
\begin{equation}  \label{13}
S(q,I,\vec{\epsilon}
(\Omega))=\frac{q^2}{2}\frac{\dot{\rho}}{\rho}+
I{\rm arcsin}\Bigg[\frac{q}{\sqrt{2I\rho ^2}}\Bigg]+
         \frac{q\sqrt{2I\rho ^2-q^2}}{2\rho ^2}~, 
\end{equation}
where we have put to zero the constant of integration. Then
\begin{equation} \label{14}
\theta=\frac{\partial S}{\partial I}={\rm arcsin}
\Big(\frac{q}{\sqrt{2I\rho ^2}}\Big)~.
\end{equation}
Now, the canonical variables are 
\begin{equation} \label{15}
q_1=\rho \sqrt{2I}\sin \theta ~,\qquad
p_1=\frac{\sqrt{2I}}{\rho}\Big(\cos \theta+
\dot{\rho}\rho\sin \theta\Big)~.
\end{equation}
The dynamical angle is
\begin{equation} \label{16}
\Delta \theta ^{d}=
\int _{\Omega _{0}}^{\Omega}
\langle\frac{\partial H_{\rm{new}}}{\partial I}\rangle
d\Omega ^{'}=
\int _{\Omega _{0}}^{\Omega}\Bigg[\frac{1}{\rho ^2}-\frac{\rho ^2}{2}
\frac{d}{d\Omega ^{'}}\Big(\frac{\dot{\rho}}{\rho}\Big)\Bigg]d\Omega ^{'}~,
\end{equation}
while the geometrical angle (generalized Hannay angle) is 
\begin{equation}  \label{17}
\Delta \theta ^{g}=\frac{1}{2}\int _{\Omega _0}^{\Omega}
\Bigg[\frac{d}{d\Omega ^{'}}
(\dot{\rho}\rho)-2\dot{\rho}^2\Bigg]
d\Omega ^{'}~.
\end{equation}
The sum of $\Delta \theta ^{d}$ and $\Delta \theta ^{g}$ is the total change 
of angle (Lewis' angle):
\begin{equation}  \label{16b}
\Delta \theta ^{t} =\int _{\Omega _{0}}^{\Omega}\frac{1}{\rho ^2}
d\Omega ^{'}~.
\end{equation}
Plots of the angular quantities (24-26) for $\kappa=1$ are displayed
in figs.~2,3, and 4, respectively. For $\kappa=-1$
we've got similar plots. 

\vskip 0ex
\centerline{
\epsfxsize=180pt
\epsfbox{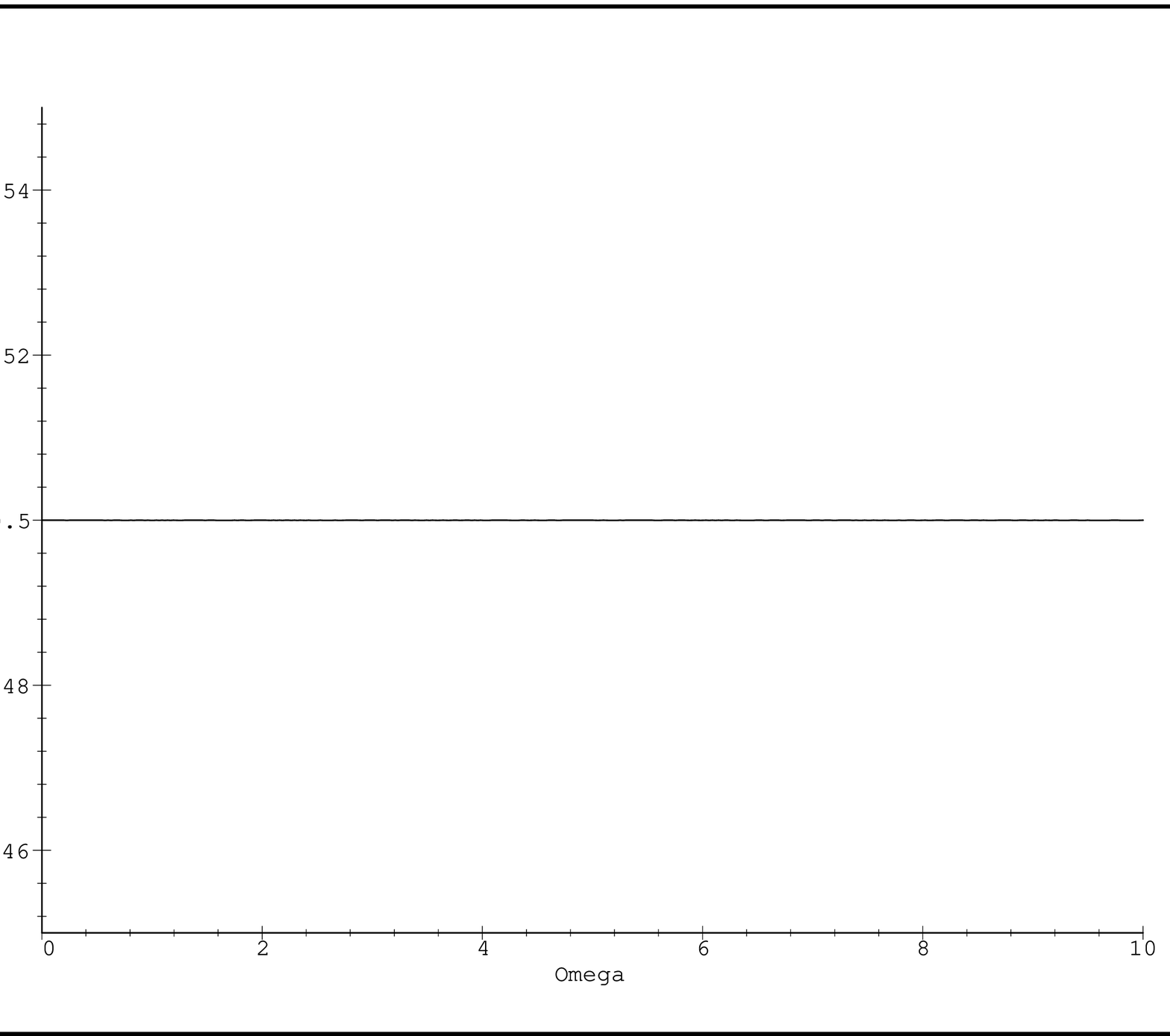}}
\vskip 0.5ex

\centerline{\footnotesize fig.~1: The Ermakov-Lewis invariant for 
$Q=0$, $\kappa =\pm1$, $h=1$.}

\vskip 0ex
\centerline{
\epsfxsize=180pt
\epsfbox{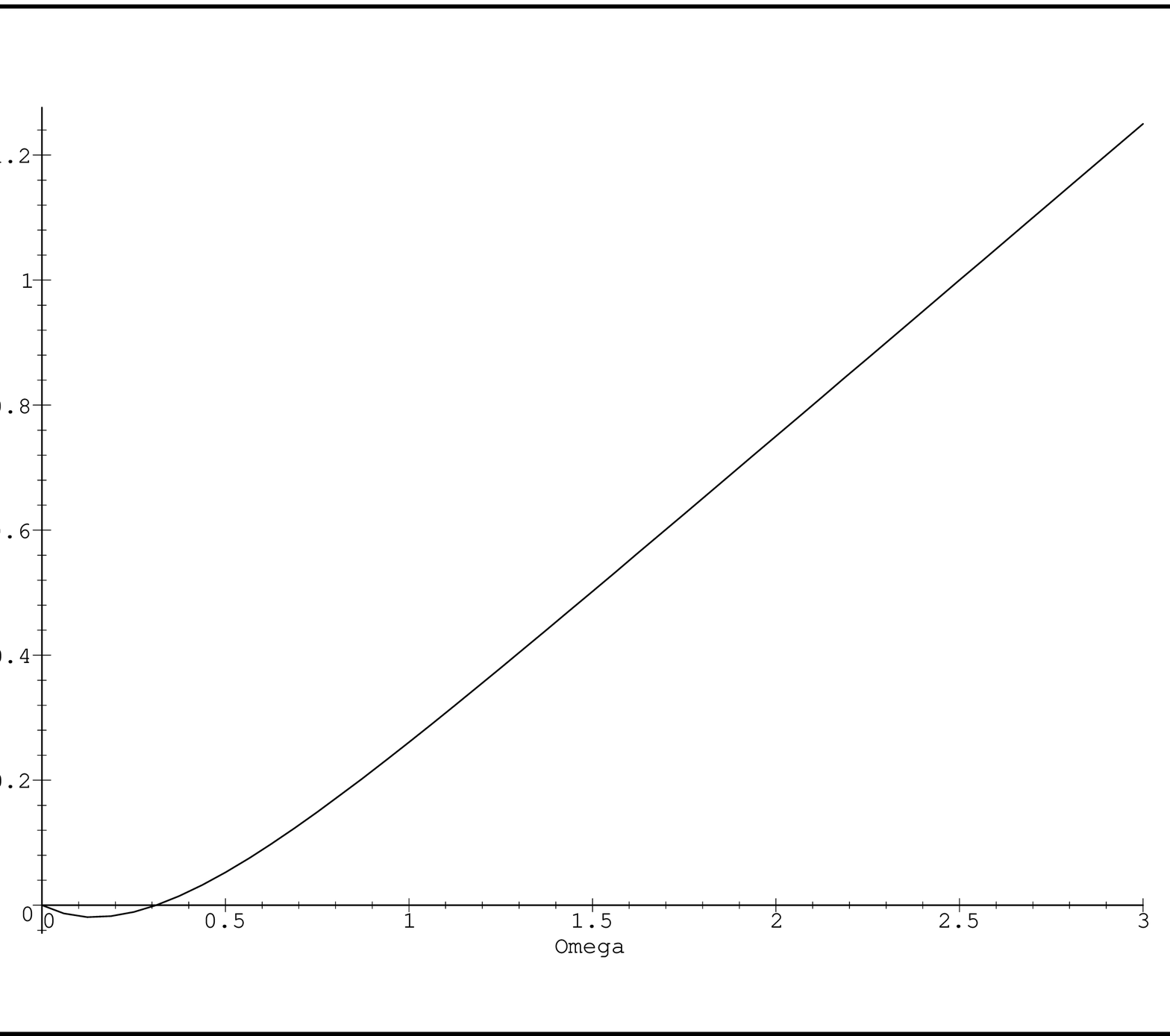}}
\vskip 0ex


\centerline{\footnotesize fig.~2: The dynamical angle as a function of 
$\Omega$.} 

\vskip 1ex
\centerline{
\epsfxsize=180pt
\epsfbox{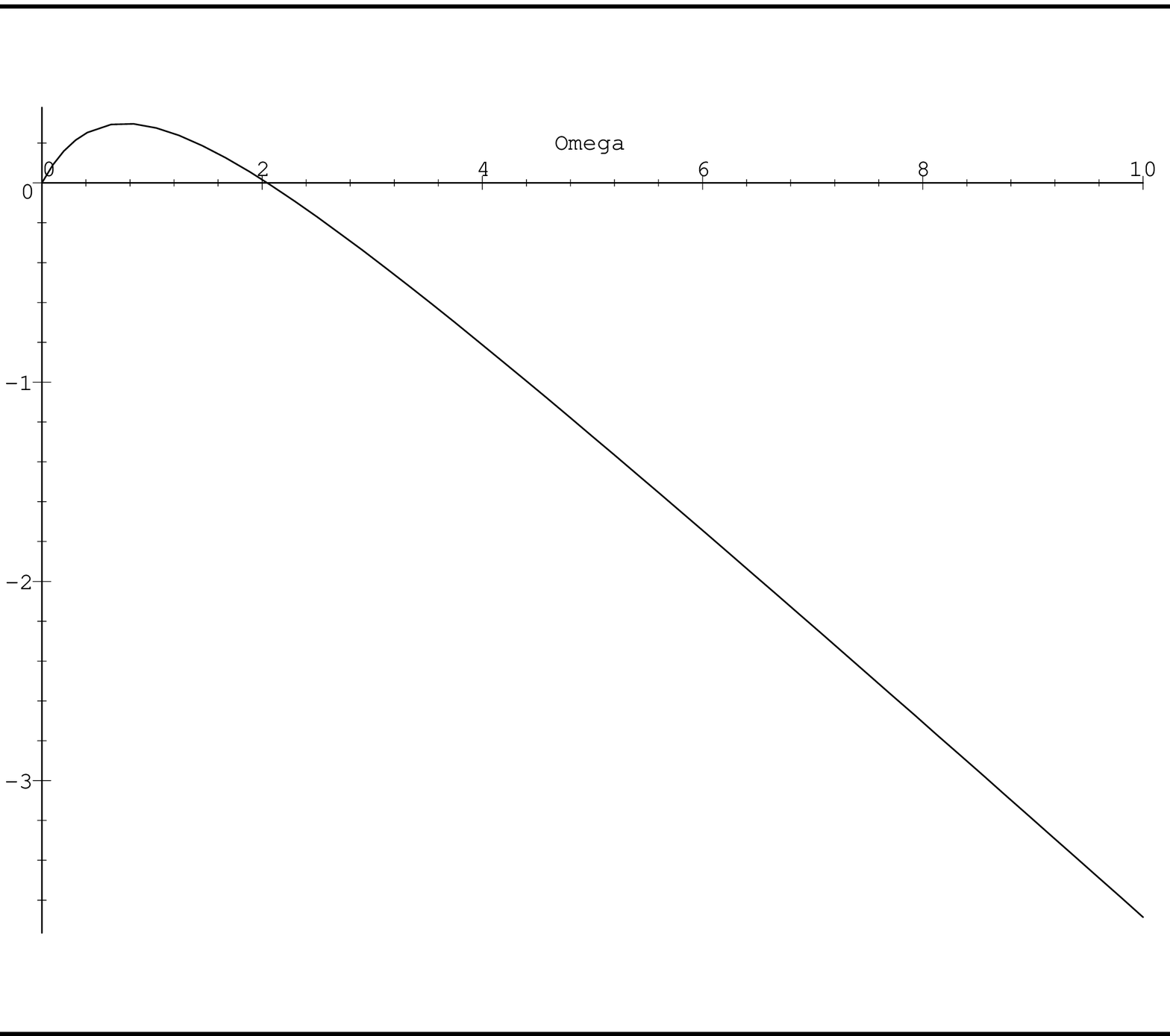}}
\vskip 1ex


\centerline{\footnotesize fig.~3: The geometrical angle as a function
of $\Omega$.} 

\vskip 1ex
\centerline{
\epsfxsize=180pt
\epsfbox{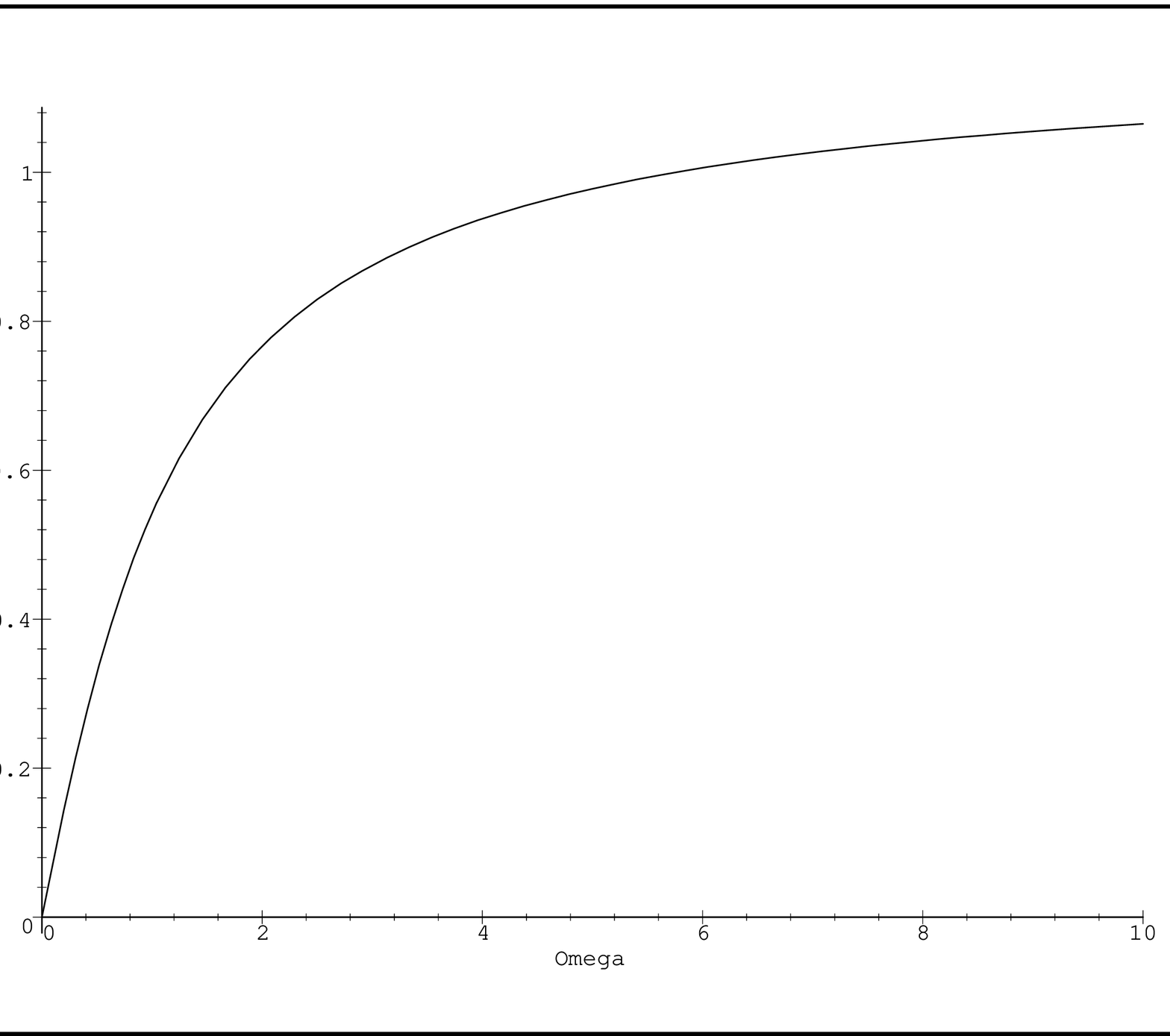}}
\vskip 1ex


\centerline{\footnotesize fig.~4: The total angle as a function of $\Omega$.}

\bigskip


\bigskip
\bigskip

\underline{EFRW universes for $Q\neq 0$.}

\bigskip

\noindent
We now apply the Ermakov procedure to the EFRW oscillators
when $Q\neq 0$. These results have been reported at the Third Workshop
of the Mexican Division of Gravitation and Mathematical Physics of the Mexican
Physical Society \cite{taller}.

It can be shown that the WDW equation for EFRW universes with $Q$ taken as a 
free parameter is 
\begin{equation} \label{1}
\frac{d^2\Psi}{d\Omega^2}+Q\frac{d\Psi}{d\Omega}-\kappa e^{-4\Omega}\Psi
(\Omega) =0~,
\end{equation}
where, as previously, $\Omega$ is Misner's time and $\kappa$ is the curvature
index of the FRW universe; 
$\kappa =1,0,-1$ for closed, flat, and open universes, respectively.
For $\kappa=\pm 1$ the general solution can be expressed in terms of  
Bessel functions
\begin{equation} \label{2}
\Psi _{\alpha}^{+}(\Omega) =
e^{-2\alpha\Omega}\left(C_1I_{\alpha}(\frac{1}{2}e^{-2\Omega})+
C_2K_{\alpha}(\frac{1}{2}e^{-2\Omega})\right)
\end{equation}
and
\begin{equation} \label{3}
\Psi _{\alpha}^{-}(\Omega) =
e^{-2\alpha\Omega}\left(C_1J_{\alpha}(\frac{1}{2}e^{-2\Omega})+
C_2Y_{\alpha}(\frac{1}{2}e^{-2\Omega})\right),
\end{equation}
respectively, where $\alpha =Q/4$. The case $\kappa =0$ does not correspond to a
parametric oscillator and will not be considered here. The eq.~(29) can be 
turned into the canonical equations for a classical point particle of mass
$M=e^{Q\Omega}$, generalized coordinate $q=\Psi$ and moment 
$p=e^{Q\Omega}\dot{\Psi}$ (i.e., of velocity $v=\dot{\Psi}$). Again, 
identifying Misner's time $\Omega$ with the classical Hamiltonian time 
we obtain the equations of motion
\begin{eqnarray}
\dot{q}\equiv\frac{dq}{d\Omega}&=&e^{-Q\Omega}p~\\   
\dot{p}\equiv
\frac{dp}{d\Omega}&=&\kappa e^{(Q-4)\Omega}q~.
\end{eqnarray}
as derived from the time-dependent Hamiltonian:
\begin{equation} \label{4}
H_{\rm cl}(\Omega)=
e^{-Q\Omega}\frac{p^2}{2}-\kappa e^{(Q-4)
\Omega}\frac{q^2}{2}~.
\end{equation}
The Ermakov invariant $I(\Omega)$ can be built algebraically to be a constant 
of motion. The result is 
\begin{equation} \label{6}
{\cal I}(\Omega)=
(\rho ^2)\cdot \frac{p^2}{2}-(e^{Q\Omega}\rho\dot{\rho})\cdot pq+
(e^{2Q\Omega}\dot{\rho} ^2 +\frac{1}{\rho ^2})\cdot\frac{q^2}{2}~, 
\end{equation}
where $\rho$ is the solution of Pinney's equation
$
\ddot{\rho}+Q\dot{\rho}-\kappa e^{-4\Omega}\rho=\frac{e^{-2Q\Omega}}
{\rho ^3}$.
In terms of $\rho _{\pm}(\Omega)$ the Ermakov invariant for this class of 
EFRW universes reads
\begin{equation} \label{10}
I _{\rm EFRW}^{\pm}=\frac{(\rho _{\pm} p-e^{Q\Omega}\dot{\rho} _{\pm}
q)^2}{2}
+\frac{q^2}{2\rho ^2_{\pm}}=
\frac{e^{2Q\Omega}}{2}\left(
\rho _{\pm} \dot{\Psi} _{\alpha}^{\pm}-\dot{\rho} _{\pm}
 \Psi _{\alpha}^{\pm}\right)^2
+\frac{1}{2}\left(\frac{\Psi _{\alpha}^{\pm}}{\rho _{\pm}}
\right)^{2}~.
\end{equation}
In the calculation of $I _{\rm EFRW}^{\pm}$ we have used linear combinations 
of Bessel functions that fulfill the initial conditions for
$\rho$ as explained in chapter 6 and which will be presented in some detail
in a separate section of the present chapter. 

Calculating again the generating function $S(q,I,\Omega)$ of the canonical 
transformations leading to the new momentum $I$, we obtain
\begin{equation}   \nonumber         
S(q,I,\Omega)=e^{Q\Omega}\frac{q^2}{2}\frac{\dot{\rho}}{\rho}+
I{\rm arcsin}\Bigg[\frac{q}{\sqrt{2I\rho ^2}}\Bigg]
         +\frac{q\sqrt{2I\rho ^2-q^2}}{2\rho ^2}~, 
\end{equation}
where the integration constant is again chosen to be zero.
The new canonical variables are $q_1=\rho \sqrt{2I}\sin \theta$ y 
$p_1=\frac{\sqrt{2I}}{\rho}\Big(\cos \theta+
e^{Q\Omega}\dot{\rho}\rho\sin \theta\Big)$.
The angular quantities are:
$\Delta \theta ^{d}=
\int _{\Omega _{0}}^{\Omega}
\langle\frac{\partial H_{\rm{new}}}{\partial {\cal I}}\rangle
d\Omega ^{'}=
\int _{0}^{\Omega}[\frac{e^{-Q\Omega '}}{\rho ^2}-
\frac{1}{2}\frac{d}{d\Omega ^{'}}
(e^{Q\Omega ^{'}}\dot{\rho}\rho)+e^{Q\Omega ^{'}}
\dot{\rho}^2]d\Omega ^{'}$,
$\Delta \theta ^{g}=\frac{1}{2}\int _{\Omega _0}^{\Omega}
[\frac{d}{d\Omega ^{'}}
(e^{Q\Omega ^{'}}\dot{\rho}\rho)-2e^{Q\Omega ^{'}}
\dot{\rho}^2]
d\Omega ^{'}$, for the dynamical and geometrical angles, respectively.
Thus, the total angle will be
$\Delta \theta =\int _{\Omega _{0}}^{\Omega}\frac{e^{-Q\Omega ^{'}}}
{\rho ^2}
d\Omega ^{'}$.
On the Misner time axis, going to $-\infty$ means going to the origin of the 
universe, whereas $\Omega _{0}=0$ means the present era. With these temporal
limits  for the cosmological evolution, one finds that the variation of the total
angle $\Delta \theta$ is basically the same as the   
Laplace transformation  of $1/\rho ^2$: $\Delta \theta=-L_{1/\rho ^{2}}(Q)$.

The plots of the invariant and of the variations of the angular quantities
are shown next, both for the closed EFRW universes as for the open ones.

\bigskip




\bigskip


\vskip 0.5ex
\centerline{
\epsfxsize=180pt
\epsfbox{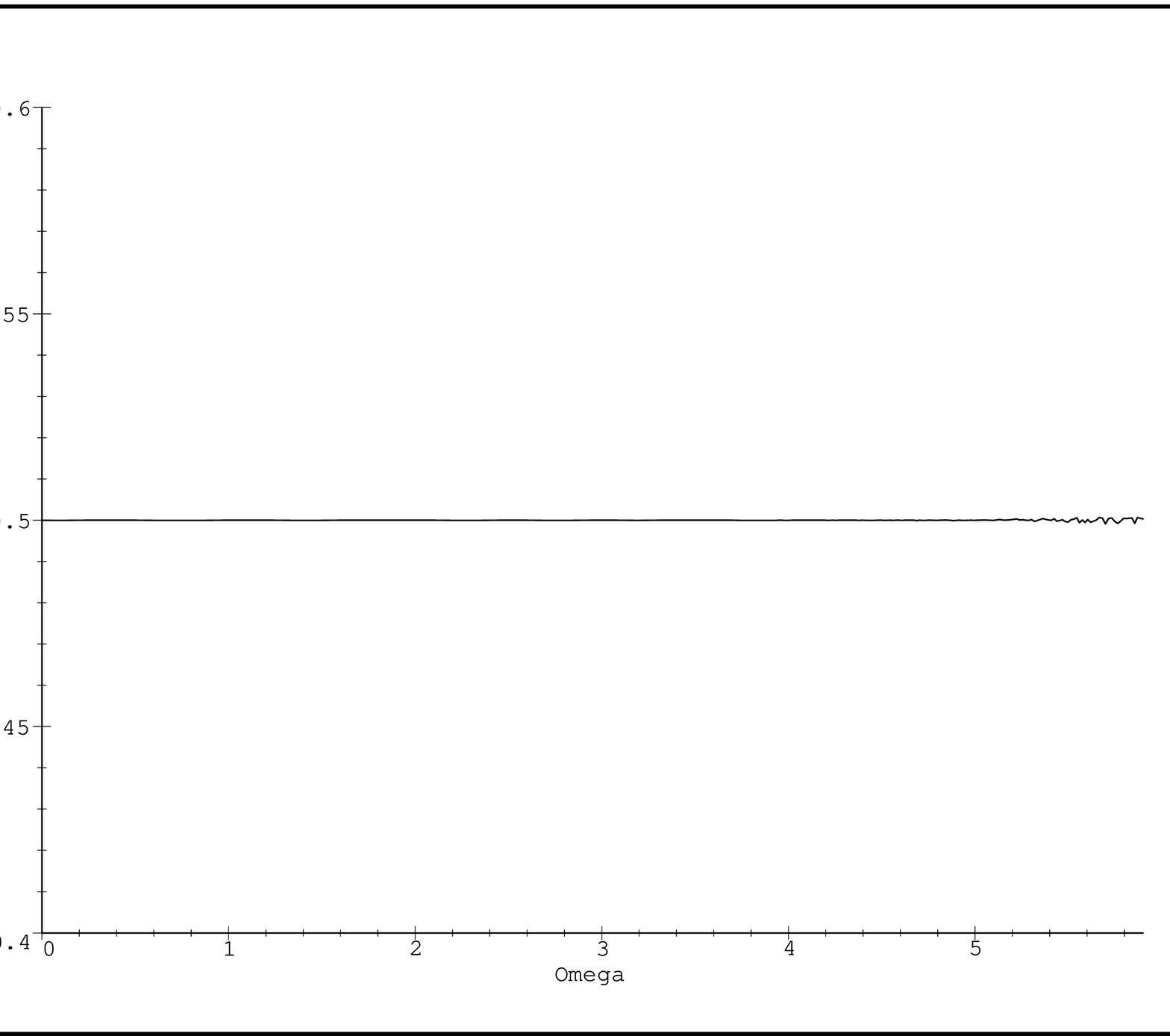}}
\vskip 1ex

\bigskip

\centerline{\footnotesize fig.~5: $I^{+} _{\rm EFRW}(\Omega)$  
for $Q=3$ and an initial singularity of unit auxiliary angular momentum.}

\vskip 0.5ex
\centerline{
\epsfxsize=180pt
\epsfbox{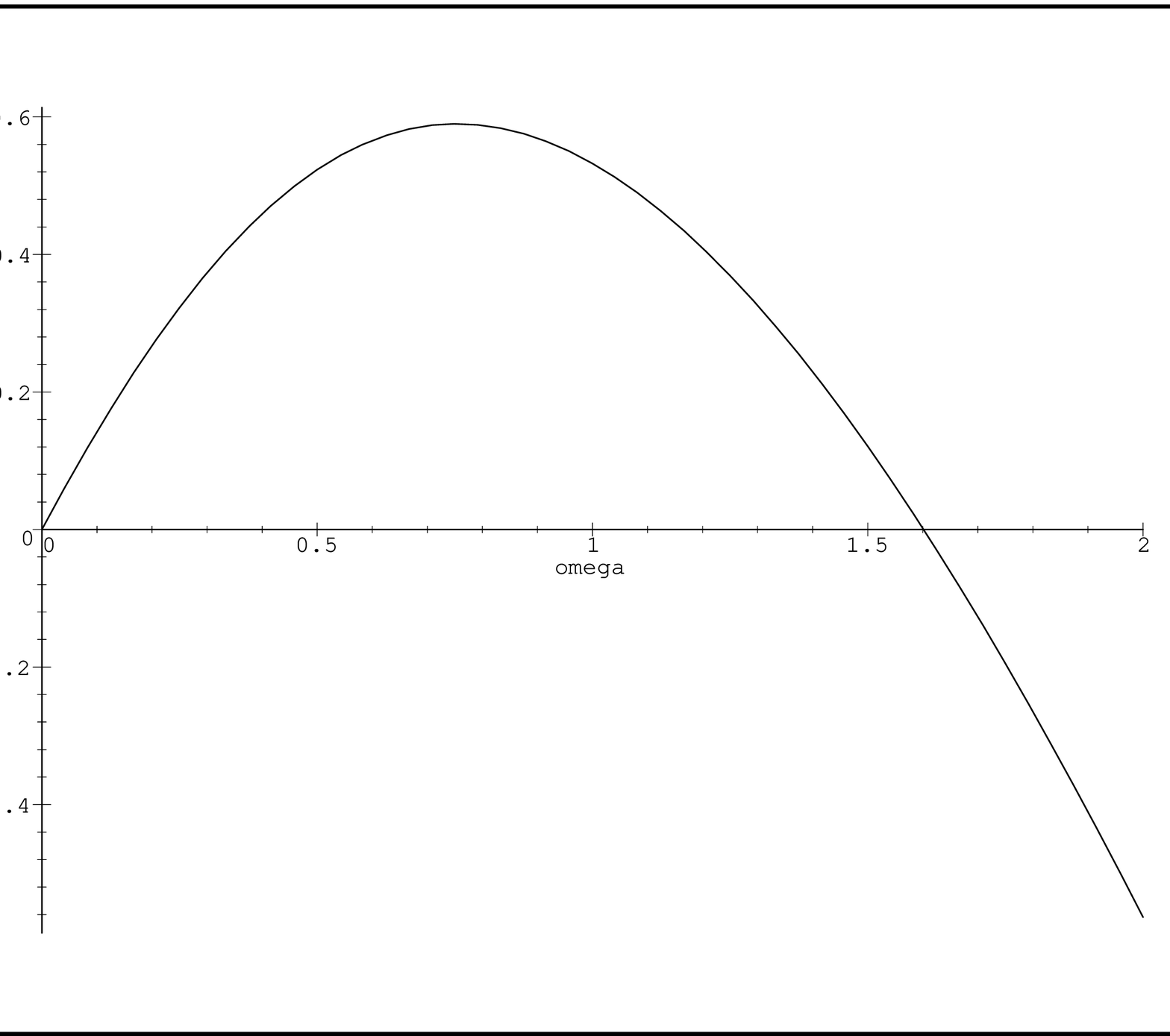}}
\vskip 1ex

\bigskip

\centerline{\footnotesize fig.~6: The dynamical angle as a function of  
$\Omega$ for a closed EFRW universe and $Q=1$.} 

\bigskip

\vskip 1ex
\centerline{
\epsfxsize=180pt
\epsfbox{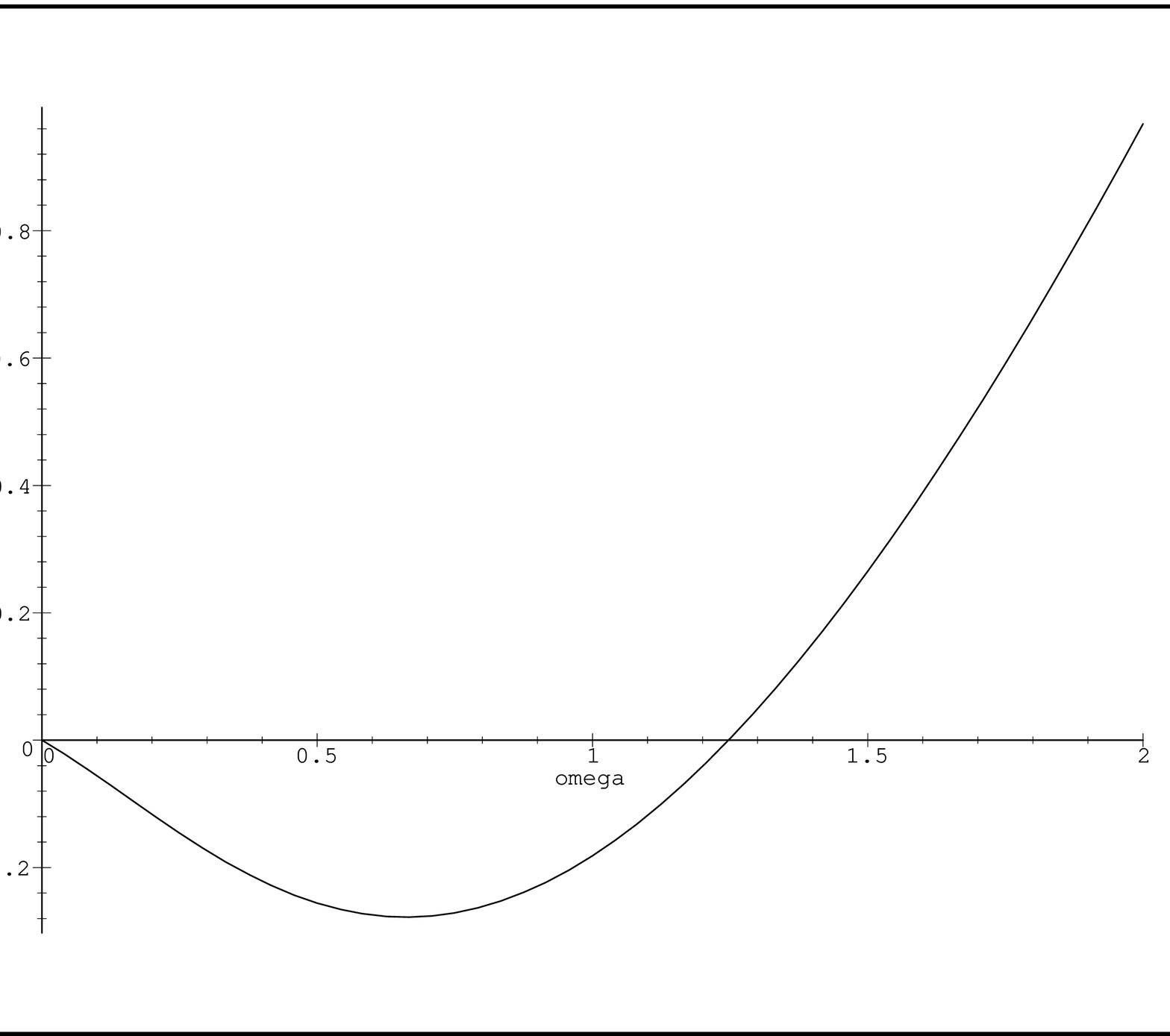}}
\vskip 1ex


\centerline{\footnotesize fig.~7: The geometrical angle for the same case.} 

\bigskip

\vskip 1ex
\centerline{
\epsfxsize=180pt
\epsfbox{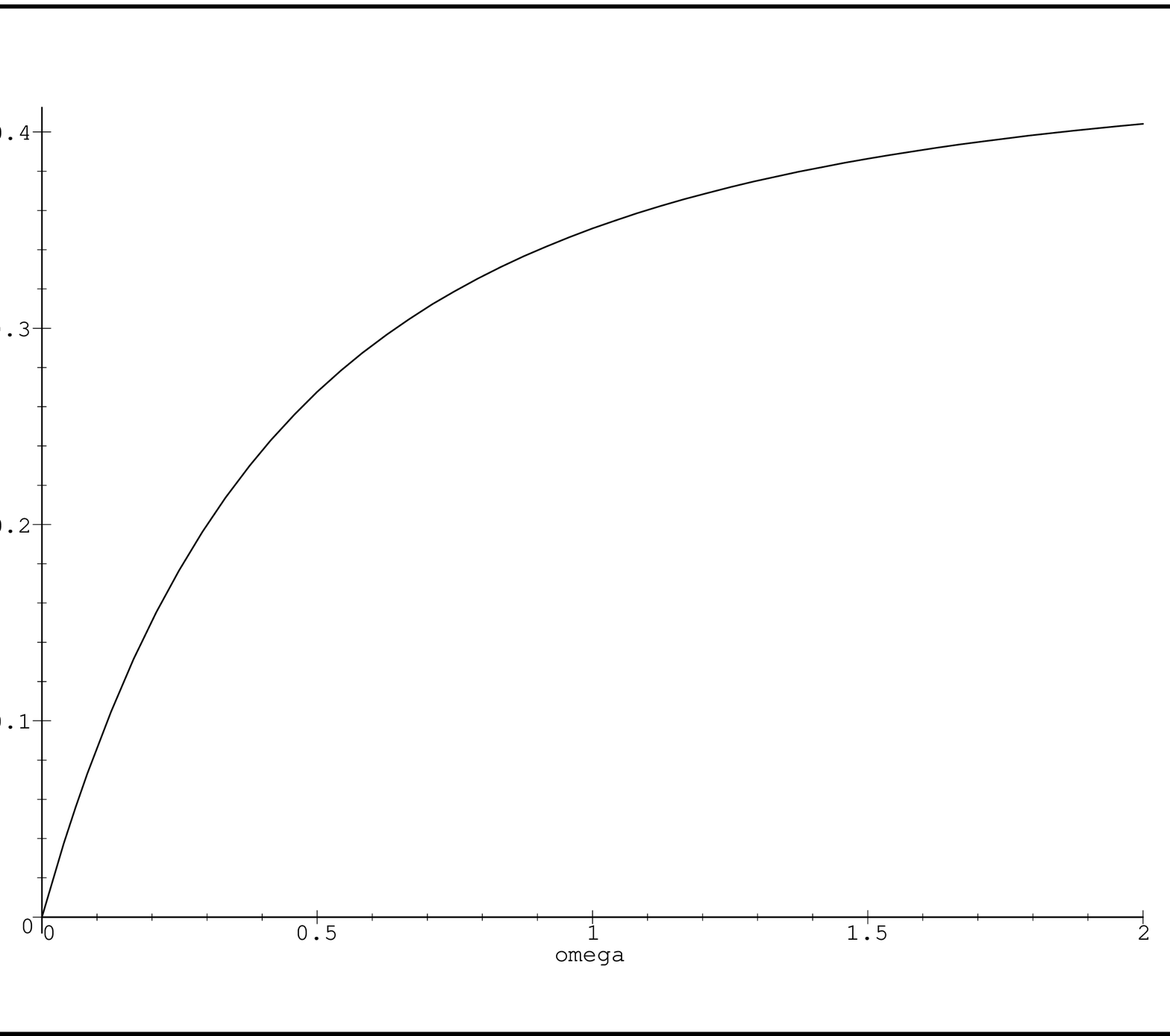}}
\vskip 1ex


\centerline{\footnotesize fig.~8: The total angle as a function of $\Omega$ 
for the same case.}

\bigskip

\vskip 0.5ex
\centerline{
\epsfxsize=180pt
\epsfbox{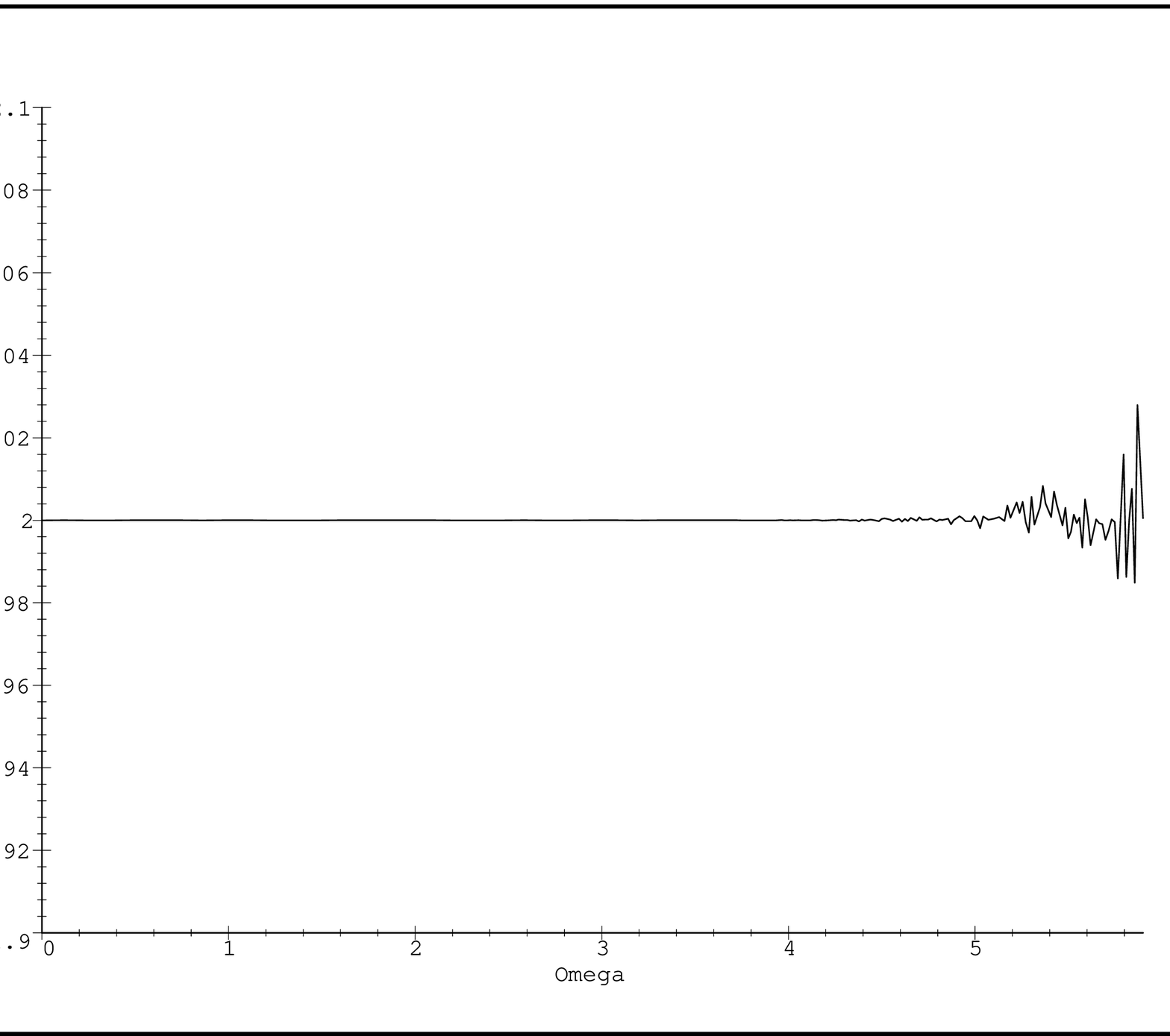}}
\vskip 1ex

\bigskip

\centerline{\footnotesize fig.~9: $I^{-} _{\rm EFRW}(\Omega)$  
with $Q=1$ for an initial singularity of auxiliary angular momentum excitation  
$h=2$.}

\vskip 1ex
\centerline{
\epsfxsize=180pt
\epsfbox{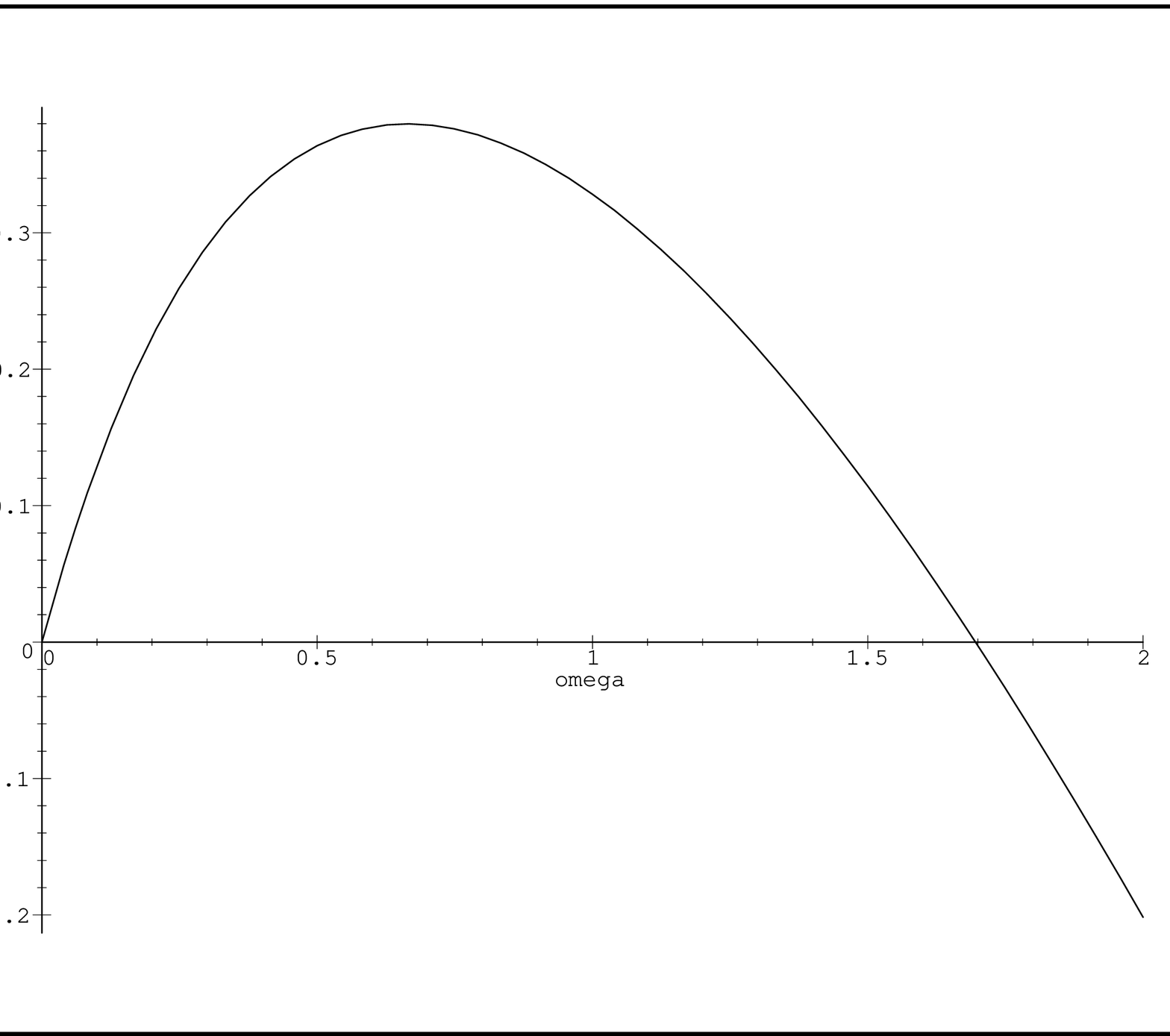}}
\vskip 1ex

\centerline{\footnotesize fig.~10: The dynamical angle as a function of $\Omega$ 
for an open EFRW universe of $Q=1$.}

\bigskip

\vskip 1ex
\centerline{
\epsfxsize=180pt
\epsfbox{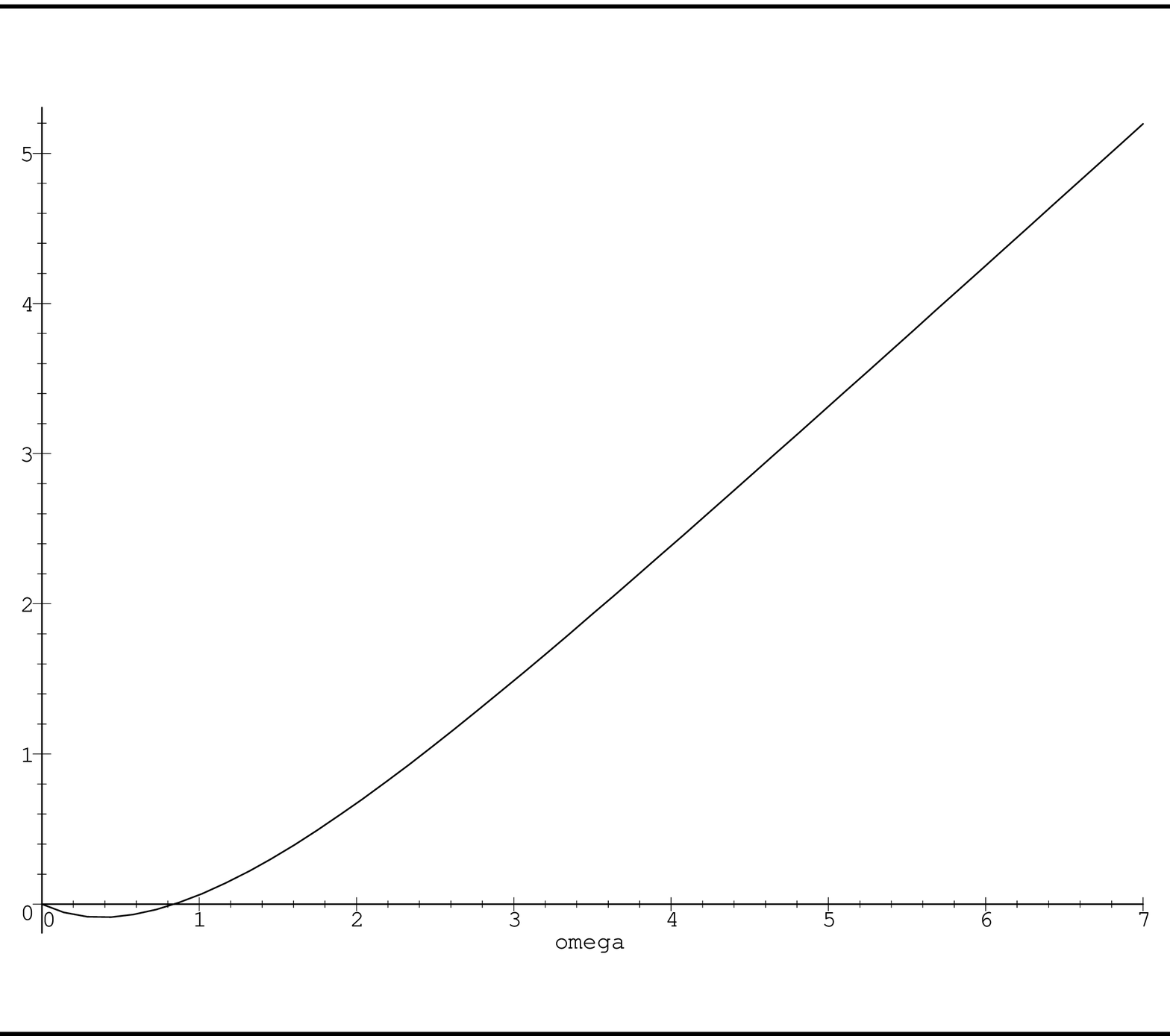}}
\vskip 1ex

\centerline{\footnotesize fig.~11: The geometrical angle
as a function of $\Omega$ for the same case.}

\bigskip

\vskip 1ex
\centerline{
\epsfxsize=180pt
\epsfbox{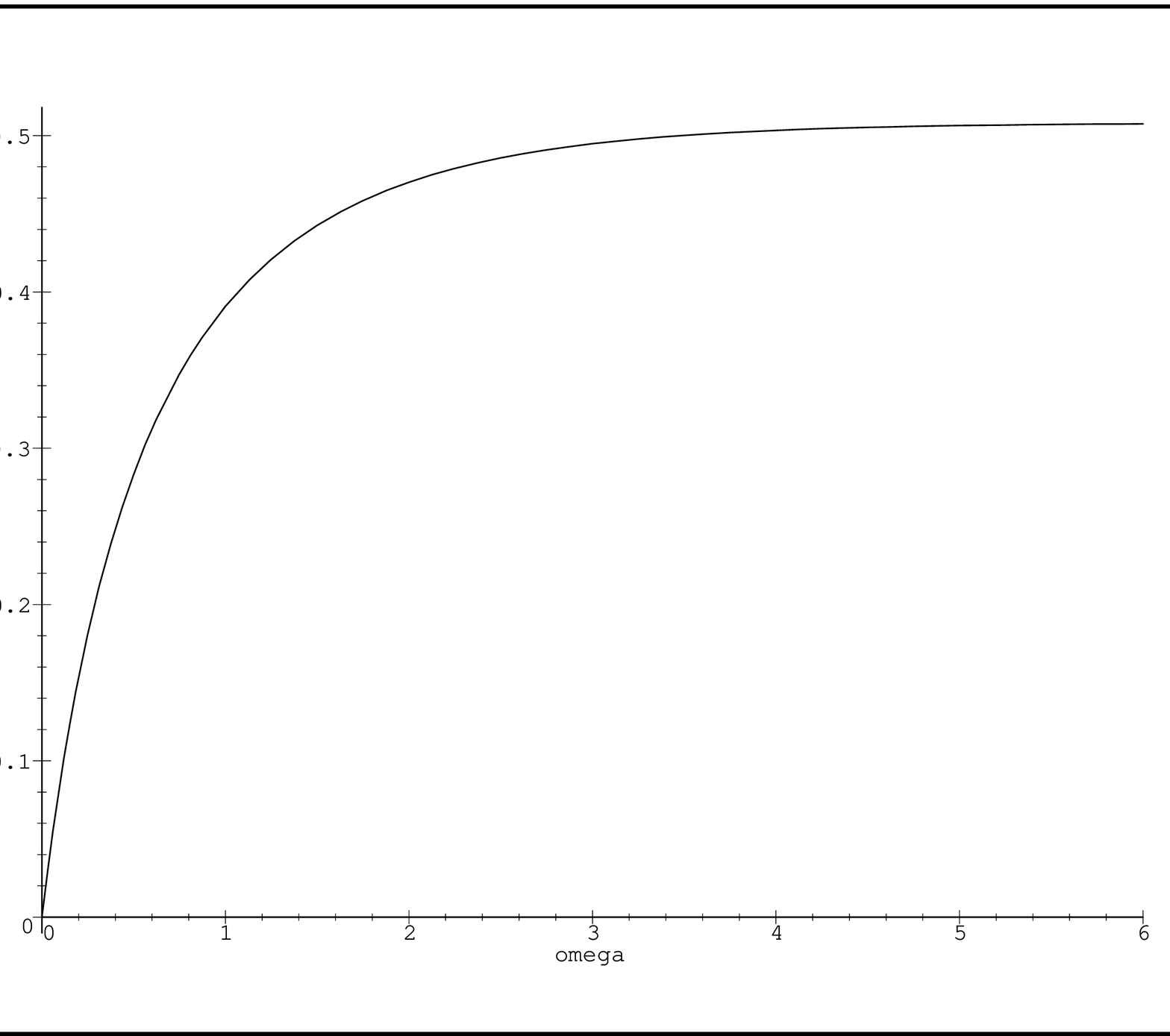}}
\vskip 1ex

\centerline{\footnotesize fig.~12: The total angle as a function of $\Omega$ 
for the same open case.}

\bigskip
\setcounter{equation}{0}
\underline{Somewhat more complicated cosmological models}

\bigskip

We sketch now the Taub  pure gravity model whose
WDW equation reads
\begin{equation}
\frac{\partial ^2\Psi}{\partial \Omega ^2}-
\frac{\partial ^2\Psi}{\partial \beta ^2}+Q\frac{\partial \Psi}
{\partial \Omega}+e^{-4\Omega}V(\beta)\Psi=0~,
\end{equation}
where $V(\beta)=\frac{1}{3}(e^{-8\beta}-4e^{-2\beta})$.
This  equation can be separated in the variables
$x_1=-4\Omega-8\beta$ and $x_2=-4\Omega -2\beta$. Thus one is led to the 
following pair of 1D differential equations for which the Ermakov procedure is 
similar to the EFRW case
\begin{equation}
\frac{d ^2 \Psi _{T1}}{d x_1^2}+\frac{Q}{12}\frac{d\Psi _{T1}}{dx_1}
+\left(\frac{\omega ^2}{4}-\frac{1}{144}e^{x_1}\right)
\Psi _{T1}=0  
\end{equation}
and
\begin{equation}
\frac{d^2 \Psi _{T2}}{d x_2^2}-\frac{Q}{3}\frac{d\Psi _{T2}}{dx_2}
+\left(\omega ^2-\frac{1}{9}e^{x_2}\right)\Psi _{T2}=0~.
\end{equation}
where $\omega /2$ is a separation constant.
The solutions are $\Psi _{T1}\equiv \Psi _{T\alpha_{1}}
=e^{(-Q/24)x_1}Z_{i\alpha _1}(ie^{x_1/2}/6)$ and
$\Psi _{T2}\equiv \Psi _{T\alpha _{2}}
=e^{(Q/6)x_2}Z_{i\alpha _2}(i2e^{x_2/2}/3)$,
respectively, where $\alpha _1=
\sqrt{\omega ^2-(Q/12)^2}$ and $\alpha _2=\sqrt{4\omega ^2-(Q/3)^2}$.

\bigskip

A more realistic case is that in which a scalar field of minimal coupling
to the FRW minisuperspace metric is included.
The WDW equation is
\begin{equation}
[\partial ^2_{\Omega}+Q\partial _{\Omega}-\partial ^{2}_{\phi}-\kappa
e^{-4\Omega}+m^2e^{-6\Omega}\phi ^2]\Psi (\Omega ,\phi)=0~,
\end{equation}
and can be written as a 
Schroedinger equation for a two-component wave function  
(see \cite{mosta}). This allows to think of squuezed cosmological states in the 
Ermakov framework \cite{ped}. For this, we shall use the following 
factorization of the invariant ${\cal I}=\hbar(bb^{\dagger}+\frac{1}{2})$,
where
$
b          =(2\hbar )^{-1/2}
  [\frac{q}{\rho}+i(\rho p-e^{Q_{c}\Omega}\dot{\rho}q)]
$
and
$
b^{\dagger}=
(2\hbar )^{-1/2}[\frac{q}{\rho}-i(\rho p-e^{Q_{c}\Omega}
\dot{\rho}q)]
$.
$Q_{c}$ is a fixed ordering parameter. 
Consider now a Misner reference oscillator of frequency 
$\omega _{0}$ corresponding to a given cosmological epoch 
$\Omega _{0}$ for which one can introduce the standard factorization operators 
$a          =(2\hbar \omega _{0})^{-1/2}[\omega _{0}q+ip],\quad
a^{\dagger}=(2\hbar \omega _{0})^{-1/2}[\omega _{0}q-ip]
$.
The connection between the two pairs $a$ and $b$ is
$b(\Omega)          =\mu(\Omega)a+\nu(\Omega)a^{\dagger}$ y
$b^{\dagger}(\Omega)=\mu ^{*}(\Omega)a^{\dagger}+\nu ^{*}
(\Omega)a^{\dagger}~, 
$
where
$\mu(\Omega)=(4\omega _{0})^{-1/2}[\rho ^{-1}-ie^{Q_{c}\Omega}
\dot{\rho}+\omega _{0}\rho]$
and
$\nu(\Omega)=(4\omega _{0})^{-1/2}[\rho ^{-1}-ie^{Q_{c}\Omega}
\dot{\rho}-\omega _{0}\rho]$ satisfy the relationship 
$|\mu(\Omega)|^{2}-|\nu(\Omega)|^{2}=1$.
The uncertainties can be calculated 
$(\Delta q)^{2}=\frac{\hbar}{2\omega _{0}}|\mu - \nu|^{2}$,
$(\Delta p)^{2}=\frac{\hbar \omega _{0}}{2}|\mu + \nu|^{2}$, and
$(\Delta q)(\Delta p)= \frac{\hbar}{2}|\mu +\nu||\mu -\nu|$ showing that
in general these Ermakov states are not of minimum uncertainty \cite{ped}.

\bigskip
\setcounter{equation}{0}
\underline{The way one should do the linear combinations for the solutions}

\underline{of the linear differential equations.}
\newline

As it has been shown, in order to solve Pinney's equation one should first find 
the solutions to the equations of motion. 
Since these equations are linear, we have chosen those combinations which
satisfy the initial conditions of motion. 
According to the interpretation of Eliezer and Gray, the solution of
Pinney's equation is just the amplitude of the 2D auxilliary motion. 
Therefore, two of the three quadratic terms of the solution can be seen as the
amplitudes along each of the axes, respectively. The third one is a mixed term 
(one can also eliminate it by diagonalizing the quadratic form in the square 
root).

Let $q(0)=a$ and $\dot{q}(0)=b$ be the initial conditions
for the equation of motion. 
The solution can be written as  
$x\left( t\right)=a x_{\rm 1}\left( t\right) 
+b x_{\rm 2}\left( t\right)$, and therefore the functions $x_{\rm 1}$ y 
$x_{\rm 2}$ must satisfy the conditions $x_{\rm 1}\left( 0\right) =1$, 
$\dot x_{\rm 1}\left( 0\right) =0$, $x_{\rm 2}\left( 0\right) =0$, 
$\dot x_{\rm 2}\left( 0\right) =1$. If we take $\psi_{\rm 1}$ and 
$\psi_{\rm 2}$ as a pair of linear independent solutions of the parametric
equation, then we can build the functions $x_{\rm i}$ as linear combinations
of $\psi_{\rm i}$: 
$x_{\rm i}=a_{\rm i}\psi_{ 1}+b_{\rm i}\psi_{2}$. It is clear that the linear
superpositions that satisfy the initial conditions will be:
\begin{equation} \label{c3-1}
x_{ 1}=\frac{1}{W(0)}\left[ \psi ^{'}_{2}(0)\psi _{1}(t)-
\psi ^{'}_{ 1}(0)\psi_{ 2}(t)\right]
\end{equation}
\begin{equation}\label{c3-2}
x_{ 2}=\frac{1}{W(0)}\left[ -\psi_{2}(0)\psi_{ 1}(t)+
\psi _{ 1}(0)\psi _{ 2}(t)\right]
\end{equation}
where $W(0)$ is the Wronskian of the functions $\psi _{ 1}$ and $\psi _{ 2}$ 
evaluated at zero time parameter. 
The functions $x_{\rm i}$ are the correct ones that should enter the solution of 
Pinney's equation written in the form given by Eliezer and Gray. 
In this way, we have in the case of the cosmological models that have been 
discussed:
\begin{equation} \label{c3-3}
x_{\rm 1}=\frac{\left( 2z\right)^\frac{Q}{4}}{2}
\left[{\psi_{\rm 1}}^{'}(1/2)K_{\frac{Q}{4}}(z)
-{\psi_{\rm 2}}^{'}(1/2)I_{\frac{Q}{4}}(z) \right]
\end{equation}
\begin{equation} \label{c3-4}
x_{\rm 2}=\frac{\left( 2z\right)^\frac{Q}{4}}{2}
\left[K_{\frac{Q}{4}}(1/2)I_{\frac{Q}{4}}(z)
-I_{\frac{Q}{4}}(1/2)K_{\frac{Q}{4}}(z) \right]
\end{equation}
where 
$$
z=\frac{1}{2}e^{-2\Omega}~,
$$
$$
{\psi_{\rm 1}}^{'}(1/2)=
-\left[\frac{Q}{2}I_{\frac{Q}{4}}(1/2)+{I_{\frac{Q}{4}}}^{'}(1/2)\right]~,
$$  
$$
{\psi_{\rm 2}}^{'}(1/2)=-\left[\frac{Q}{2}K_{\frac{Q}{4}}(1/2)
+{K_{\frac{Q}{4}}}^{'}(1/2)\right]~,
$$ 
for closed EFRW, and similarly for the open EFRW models.

The superposition coefficients we worked with are of the form
$a_{+}=N_K(1/2)/D_{+}(1/2)$, $b_{+}=N_I(1/2)/D_{+}(1/2)$,
$c_{+}=-K(1/2)/D_{+}(1/2)$, $d_{+}=I(1/2)/D_{+}(1/2)$, where 
$N_K(1/2)=K_{\frac{Q}{4}+1}(1/2)-QK_{\frac{Q}{4}}(1/2)$,
$N_I(1/2)=I_{\frac{Q}{4}+1}(1/2)+QK_{\frac{Q}{4}}(1/2)$, and
$D_{+}(1/2)=I_{\frac{Q}{4}+1}(1/2)K_{\frac{Q}{4}}(1/2)+
K_{\frac{Q}{4}+1}(1/2)I_{\frac{Q}{4}}(1/2)$ for the closed EFRW case;
$a_{-}=-N_Y(1/2)/D_{-}(1/2)$, $b_{-}=N_J(1/2)/D_{-}(1/2)$,
$c_{-}=Y(1/2)/D_{-}(1/2)$, $d_{-}=-J(1/2)/D_{-}(1/2)$, where 
$N_Y(1/2)=Y_{\frac{Q}{4}+1}(1/2)-QY_{\frac{Q}{4}}(1/2)$,
$N_J(1/2)=J_{\frac{Q}{4}+1}(1/2)+QJ_{\frac{Q}{4}}(1/2)$, and
$D_{-}(1/2)=J_{\frac{Q}{4}+1}(1/2)Y_{\frac{Q}{4}}(1/2)-
Y_{\frac{Q}{4}+1}(1/2)J_{\frac{Q}{4}}(1/2)$ for the open EFRW case.

\bigskip
\setcounter{equation}{0}
\section*{ 11. Application to physical optics.}

In order to study the Ermakov procedure within physical optics,
our starting point will be  the 1D Helmholtz equation in the form given by 
Goyal {\em et al} \cite{Goy} and Delgado {\em et al} \cite{Del}
\begin{equation} \label{o1}
\frac{d^2\psi}{dx^2}+\lambda \phi(x)\psi (x) =0~,
\end{equation}
that is, as a Sturm-Liouville equation for the set of eigenvalues 
$\lambda \in R$ defining the Helmholtz spectrum within a closed given interval 
[a,b] on the real line, where the nontrivial function $\psi$ turns to zero at the 
end points 
(Dirichlet boundary conditions). Eq.~(1) occurs, for example, 
in the case of the transversal electric modes 
(TE) propagating in waveguides that have a continuously varying refractive index  
in the $x$ direction but are independent of $y$ and $z$. Similar problems 
in acoustics can be treated along the same lines. 
The transformation of eq.~(1) into the canonical equations
of motion of a classical pointparticle is performed as follows.
Let $\psi (x)$ by any real solution of eq.~(1). 
Define $x=t$, $\psi =q$, and $\psi ^{'}=p$; then, eq.~(1) turns into
\begin{eqnarray}
\frac{dq}{dt}&=&p~\\
\frac{dp}{dt}&=&-\lambda\phi(t)q~,
\end{eqnarray}
with the boundary conditions $q(a)=q(b)=0$. The corresponding classical 
Hamiltonian 
\begin{equation} \label{o4}
H(t)=\frac{p^2}{2}+\lambda\phi(t)\frac{q^2}{2}~.
\end{equation}
is similar to the previous cosmological case of $Q=0$, if one identifies 
$\lambda=-\kappa$ and $\phi =e^{-4\Omega}$. 
The procedure to find the Ermakov invariant follows step by step the cosmological 
case. In the phase space algebra we can write the invariant as
\begin{equation} \label{o5}
I=\sum _{r}\mu _{r}(t)T_{r}~,
\end{equation}
and applying 
\begin{equation} \label{6}
\frac{\partial I}{\partial t}=-\{I,H\}~,
\end{equation}
we get the system of equations for the coefficients $\mu _{r}(t)$
\begin{eqnarray} \nonumber
\dot{\mu} _1&=&-2\mu _2 \\
\dot{\mu} _2&=&\lambda \phi(t)\mu _1-\mu _3\\
\dot{\mu} _3&=&2\lambda \phi(t)\mu _2~.    \nonumber
\end{eqnarray}
The solutions can be written in the conventional form by choosing 
$\mu _1=\rho ^2$, that gives $\mu _2=-\rho \dot{\rho}$ and 
$\mu _3=\dot{\rho} ^2+
\frac{1}{\rho ^2}$, where $\rho$ is a solution of the Pinney's equation of the 
form:
$
\ddot{\rho}+\lambda \phi(t)\rho=\frac{1}{\rho ^3},
$
with the Ermakov invariant of the well-known form
$ 
I=\frac{(\rho p-\dot{\rho}q)^2}{2}+\frac{q^2}{2\rho ^2}~.
$ 
Next, we calculate the generating function of the canonical transformation
for which $I$ is the new momentum
\begin{equation} \label{o11}
S(q,I,\vec{\mu}(t))=\int ^{q}dq^{'}p(q^{'},I,\vec{\mu}(t))~.
\end{equation}
Thus,
\begin{equation} \label{o12}
S(q,I,\vec{\mu}(t))=\frac{q^2}{2}\frac{\dot{\rho}}{\rho}+
I{\rm arcsin}\Bigg[\frac{q}{\sqrt{2I\rho ^2-q^2}}\Bigg]+
\frac{q\sqrt{2I\rho ^2-q^2}}{2\rho ^2}~, 
\end{equation}
where we have put to zero the integration constant.
In this way we get
\begin{equation} \label{o13}
\theta=\frac{\partial S}{\partial I}={\rm arcsin}
\Big(\frac{q}{\sqrt{2I\rho ^2-q^2}}\Big)~.
\end{equation}
The new canonical variables are $q_1=\rho \sqrt{2I}\sin \theta$ and $
p_1=\frac{\sqrt{2I}}{\rho}\Big(\cos \theta+\dot{\rho}\rho\sin \theta\Big)$.
The dynamical angle is given by
\begin{equation} \label{16}
\Delta \theta ^{d}=\int _{t_0}^{t}\Bigg[\frac{1}{\rho ^2}-\frac{\rho ^2}{2}
\frac{d}{dt^{'}}\Big(\frac{\dot{\rho}}{\rho}\Big)\Bigg]dt^{'}
\end{equation}
whereas the geometrical angle is
\begin{equation}  \label{o17}
\Delta \theta ^{g}=\frac{1}{2}\int _{t_0}^{t}
\Bigg[(\ddot{\rho}\rho)-\dot{\rho}^2\Bigg] dt^{'}~.
\end{equation}
For periodic parameters $\vec{\mu}(t)$, with all the components of the same
period $T$, the geometric angle is known as the nonadiabatic Hannay angle
\cite{book} that can be written as a function of $\rho$:
\begin{equation}  \label{o17b}
\Delta \theta ^{g}_{H}=-\oint _{C}\dot{\rho}d\rho~.
\end{equation}

\noindent
Now, in order to proceed with the quantization of the Ermakov 
problem, we turn $q$ 
and $p$ into operators, $\hat{q}$ y 
$\hat{p}=-i\hbar\frac{\partial}{\partial q}$, but keeping the auxiliary function
$\rho$ as a real number. The Ermakov invariant is now a Hermitian constant 
operator  
\begin{equation}  \label{o18}
\frac{d\hat{I}}{dt}=\frac{\partial\hat{I}}{\partial t}
+\frac{1}{i\hbar}[\hat{I},\hat{H}]=0
\end{equation}
and the time-dependent Schr\"odinger equation for the Helmholtz Hamiltonian is
\begin{equation}  \label{o19}
i\hbar\frac{\partial}{\partial t}|\psi (\hat{q},t)\rangle=
\frac{1}{2}(\hat{p}^2+\lambda\phi(t)\hat{q}^2)|\psi(\hat{q},t)\rangle ~.
\end{equation}

\noindent
The problem now is to find the eigenvalues of $\hat{I}$
\begin{equation}  \label{o20}
\hat{I}|\psi _{n}(\hat{q},t\rangle=\kappa _{n}|\psi _{n}(\hat{q},t)\rangle
\end{equation}
and also to write the explicit form  of the general solution of eq.~(\ref{o19})
\begin{equation}  \label{o21}
\psi(\hat{q},t)=\sum _{n}C_{n}e^{i\alpha _{n}(t)}\psi _{n}(\hat{q},t)
\end{equation}
where $C_{n}$ are superposition constants, $\psi _{n}$ are (orthonormalized)
eigenfunctions of $\hat{I}$, and the phases $\alpha _{n}(t)$ are the Lewis 
phases \cite{mor,Maa} that can be found from the equation 
\begin{equation}  \label{o22}
\hbar \frac{d\alpha _{n}(t)}{dt}=\langle \psi _{n}|i\hbar
\frac{\partial}{\partial t}-\hat{H}|\psi _{n}\rangle~.
\end{equation}
The crucial point in the Ermakov quantum problem 
is to perform a unitary transformation in such a way as to get time-independent
eigenvalues for the new Ermakov invariant 
$\hat{I}^{'}=\hat{U}\hat{I}\hat{U}^{\dagger}$.
It is easy to obtain the required unitary transformation: 
$\hat{U}=\exp [-\frac{i}{\hbar}\frac{\dot{\rho}}{\rho}\frac{\hat{q}^2}{2}]$. The
new invariant will be $\hat{I}^{'}=\frac{\rho ^{2}\hat{p}^2}{2}+
\frac{\hat{q}^{2}}{2\rho ^{2}}$. The eigenfunctions are 
$\propto e^{-\frac{\theta ^2}{2\hbar}}H_{n}(\theta/\sqrt{\hbar})$, 
where $H_{n}$ 
are the Hermite polynomials, $\theta=\frac{q}{\rho}$, and the eigenvalues
are 
$\kappa _{n}=\hbar(n+\frac{1}{2})$. Thus, one can write the eigenfunctions 
$\psi _{n}$ as follows
\begin{equation}  \label{o23}
\psi _{n}\propto \frac{1}{\rho ^{\frac{1}{2}}}
\exp \Big(\frac{1}{2}\frac{i}{\hbar}
\frac{\dot{\rho}}{\rho}q^2\Big)\exp\Big(-\frac{q^2}{2\hbar \rho ^2}\Big)
H_{n}\Big(\frac{1}{\sqrt{\hbar}}\frac{q}{\rho}\Big)~.
\end{equation}
The factor $1/\rho ^{1/2}$ has been introduced for normalization reasons.
Using these functions and doing simple calculations one can find the geometrical 
phase
\begin{equation}  \label{o23b}
\alpha _{n}^{g}=-\frac{1}{2}(n+\frac{1}{2})\int _{t_0}^{t}
\Bigg[(\ddot{\rho}\rho)-\dot{\rho}^2\Bigg] dt^{'}~.
\end{equation}
The cyclic (nonadiabatic) Berry's phase \cite{book} is
\begin{equation}  \label{o23c}
\alpha _{B,n}^{g}=(n+\frac{1}{2})\oint _{C}\dot{\rho}d\rho~.
\end{equation}

\noindent
The results obviously depend on the explicit form of $\rho$ which in turn 
depends on the explicit form of $\phi$.

One can find that a good adiabatic parameter is the inverse of the square root
of the Helmoltz eigenvalues, 
$\frac{1}{\sqrt{\lambda}}$, with a slow ``time" variable 
$\tau=\frac{1}{\sqrt{\lambda}} t$. The adiabatic approximation has been studied 
in detail by Lewis \cite{L68}. If the Helmholtz Hamiltonian is written down as 
\begin{equation} \label{o24}
H(t)=\frac{\sqrt{\lambda}}{2}[p^2+\phi(t)q^2]~,
\end{equation}
then Pinney's equation is  
\begin{equation} \label{o25}
\frac{1}{\lambda}\ddot{\rho}+\phi(t)\rho=\frac{1}{\rho ^3}~,
\end{equation}
while the Ermakov invariant becomes a 
$1/\sqrt{\lambda}$-dependent function
\begin{equation} \label{o26}
I(1/\sqrt{\lambda})=
\frac{(\rho p-\dot{\rho}q/\sqrt{\lambda})^2}{2}+\frac{q^2}{2\rho ^2}~.
\end{equation}
In the adiabatic approximation,
Lewis \cite{L68} obtained the general Pinney solution in terms of the linear
independent solutions $f$ and $g$ of the equation of motion
$\frac{1}{\lambda}
\ddot{q}+\Omega ^2(t)q=0$ for the classical oscillator 
(see eq. (45) in \cite{L68}). Among the examples given by Lewis, it is  
$\Omega(t)=bt^{m/2}$, $m\neq-2$, $b={\rm constant}$ which is directly related to
a realistic dielectric of a waveguide since it corresponds to a power-law 
index profile ($n(x)\propto x^{m/2}$).
For this case, Lewis obtained a simple formula for  
$\rho$ of $O(1)$ order in $1/\sqrt{\lambda}$
\begin{equation}   \label{o27}
\rho _{m}=\gamma _1\Bigg[\frac{\gamma _2\pi\sqrt{\lambda}}
{(m+2)}\Bigg]^{\frac{1}{2}}
t^{\frac{1}{2}}[H_{\beta}^{(1)}(y)H_{\beta}^{(2)}(y)]^{\frac{1}{2}}~,
\end{equation}
where $H_{\beta}^{(1)}$ and $H_{\beta}^{(2)}$ are Hankel functions
of order
$\beta =1/(m+2)$, $y=\frac{2b\sqrt{\lambda}}{(m+2)}t^{\frac{m}{2}+1}$, and 
$\gamma _1=\pm 1$, $\gamma _2=\pm 1$. An even more useful technological 
application might be the following proposal of Lewis: $m=-\frac{4n}{2n+1}$, 
$n=\pm 1,\pm 2, ...$, leading to
\begin{equation}  \label{o28}
\rho _{n}=
\gamma _1\gamma _2^{\frac{1}{2}}b ^{-\frac{1}{2}}t^{\frac{n}{2n+1}}
|G(t,1/\sqrt{\lambda})|^2~,
\end{equation}
where
\begin{equation}  \label{o29}
G(t,1/\sqrt{\lambda})=
\Bigg[\sum _{k=0}^{n}(-1)^{k}\frac{(n+k)!}{k!(n-k)!}
\Big(\frac{1/\sqrt{\lambda}}
{2ib(2n+1)}\Big)^{k}t^{-\frac{k}{(2n+1)}}\Bigg]^{\frac{1}{2}}~.
\end{equation}
One gets
$\rho$ as a polynomial in the square of the adiabatic parameter, 
i.e., $\lambda ^{-1}$, of infinite radius of convergence. The topological 
quantities (angles and phases) can be calculated by substituting the explicit 
form of Pinney 's function in the corresponding formulas. Lewis \cite{L68} 
found a recursive formula in $1/\lambda$ of order $1/\lambda ^3$ that can be used
for any type of index profile. The recurrence relationship is
\begin{equation}  \label{o30}
\rho = \rho _{0}+\rho _{1}/\lambda +\rho _{2}/\lambda ^2 +
\rho _{3}/\lambda ^3+...~,
\end{equation}
where $\rho _{0}=\Omega ^{-1/2}=\phi ^{-1/4}(x)$; for the other coefficients
$\rho _{i}$ see the appendix in \cite{L68}. The main contribution to the 
topological quantities are given by $\rho _{0}$. In the case of a power-law
index profile, the geometric angle is 
\begin{equation}  \label{o31}
\Delta \theta ^{g}=-\frac{m}{4b(m+2)}\Bigg[t^{-(\frac{m}{2}+1)} -
t_{0}^{-(\frac{m}{2}+1)}\Bigg]~,
\end{equation}
and a similar formula can be written for the geometric quantum phase.
For periodic indices, one can write the Hannay  angle and Berry's phase 
according to their cyclic integral expressions. 
Finally, we notice that the choice  
$\phi (x)=\Phi (x)+ \frac{{\rm Const}}{\psi ^3(x)}$, which corresponds to 
nonlinear waveguides, leads to more general time-dependent Hamiltonians 
that have been discussed in the Ermakov perspective by Maamache \cite{Maa}.

\noindent
We have presented in a formal way the application of the Ermakov approach
to 1D Helmholtz problems. For more detailes one can look in a recent work 
by Rosu and Romero \cite{roro}.


\newpage
\section*{12. Conclusions.}

As one could see from the examples we discussed in this work, the Ermakov-Lewis 
quadratic invariants are an important method of research for parametric 
oscillator problems. They are helpful for better understanding this widespreaded
class of phenomena with applications in many areas of physics.
One can also say that the Ermakov approach gives a connection between the linear
physics of parametric oscillators and the corresponding nonlinear physics. 

The cosmological applications of the classical Ermakov procedure we presented 
herein are based on a classical particle representation of the WDW equation for 
the EFRW models.
We also notice that the Ermakov invariant is equivalent to the
Courant-Snyder invariant of use in the accelerator physics \cite{cs}, 
allowing an analogy between the physics of beams and the cosmological 
evolution as suggested by Rosu and Socorro \cite{roso}.

We end up with a possible interpretation of the Ermakov invariant
within the empty minisuperspace cosmology.
If one performs an expansion of the invariant in a power series in the adiabatic 
parameter, the principal term which defines the adiabatic regime gives the 
number of adiabatic ``quanta" and there were authors who gave classical 
descriptions of the cosmological particle production in such terms \cite{pp}. 
On the other hand, the Eliezer-Gray interpretation as an angular momentum of the 
2D auxiliary motion allows one to say that for EFRW minisuperspace models,
the Ermakov invariant gives the number of adiabatic excitations of the 
auxiliary angular momentum with which the universe is created at the initial
singularity.  

\newpage
\setcounter{equation}{0}
\section*{Appendix A: Calculation of the integral of $I$.}

The phase space integral of $I$ in chapter 5 can be calculated from the formula 
(15a) in the paper of Lewis \cite{L68}
\begin{equation} \label{a1}
I=-\frac{1}{2\pi}\int _{0}^{2\pi} X_2\frac{\partial X_1}{\partial \varphi}
d\varphi
\end{equation}
where $X_1$ and $X_2$ represent the functional dependences of $q$ and $p$,
respectively, in 
terms of the nice variables $z_1$ and $\varphi$, which have been given by Lewis in 
the formulas (38) of the same paper as follows
\begin{equation} \label{a2}
X_1=\pm \frac{z_1}{F_1 \Omega [1+{\rm tan}^2(\varphi - F_2)]^{1/2}}
\end{equation}
and
\begin{equation} \label{a3}
X_2=\pm \frac{z_1[\epsilon \frac{d\ln \rho}{dt}+\frac{1}{\rho ^2}
{\rm tan}(\varphi -F_2)]}{F_1 \Omega [1+{\rm tan}^2(\varphi - F_2)]^{1/2}}~,
\end{equation}
where $F_1$ and $F_2$ are two arbitrary functions of time.
Thus,
\begin{equation} \label{a4}
\frac{\partial X_1}{\partial \varphi}=
\pm \frac{z_1}{F_1 \Omega} [1+{\rm tan}^2(\varphi - F_2)]^{-3/2}(-1/2)2
{\rm tan}
(\varphi -F_2){\rm sec}^2(\varphi -F_2)~.
\end{equation}
We have the following integral
\begin{equation} \label{a5}
I=\frac{z_1^2}{2\pi F_1^2\Omega ^2}\int _{0}^{2\pi}
\frac{[\epsilon \frac{d\ln \rho}{dt}+\frac{1}{\rho ^2}
{\rm tan}(\varphi -F_2)]}{[1+{\rm tan}^2(\varphi - F_2)]^{2}}
{\rm tan}(\varphi -F_2){\rm sec}^2(\varphi -F_2)d\varphi~.
\end{equation}
Now, employing
\begin{equation} \label{a6}
s={\rm tan}^2(\varphi - F_2),\quad ds=2{\rm tan}(\varphi-F_2)
{\rm sec}^2(\varphi - F_2)
d\varphi 
\end{equation}
one gets
\begin{equation} \label{a7}
I\propto \int 
\frac{\frac{\epsilon d\ln \rho}{dt}ds}{(1+s)^2}+ 
\frac{1}{\rho ^2}
\int\frac{s^{1/2}ds}{(1+s)^2}~.
\end{equation}
Therefore
\begin{equation} \label{a8}
I=\frac{z_1^2}{2\pi F_1^2\Omega ^2}\Bigg[-\epsilon \frac{d\ln \rho}{dt}
\frac{1}{(1+s)}+\frac{1}{\rho ^2}\left(-s^{-1/2}+
{\rm tan}^{-1}\sqrt{s}\right)\Bigg]~.
\end{equation}
Going back to the $\varphi$ variable and taking into account the 
corresponding $0$ and $2\pi$ limits one gets
\begin{equation} \label{a9}
I=\frac{z_1^2}{2 F_1^2\Omega ^2\rho ^2}~,
\end{equation}
which is the result obtained by Lewis. The common form of $I$ can be obtained 
by going back to the $(q,p)$ variables.

\bigskip
\bigskip

\setcounter{equation}{0}
\section*{Appendix B: Calculation of the expectation value of $\hat{H}$ in 
eigenstates of $\hat{I}$.}

From the formulas (12) in chapter 5 for the raising and lowering operators one 
gets
\begin{equation} \label{B1}
\hat{q}=\frac{\rho}{\sqrt{2}}(\hat{a}^{+}+\hat{a})~,
\end{equation}
\begin{equation} \label{B2}
\hat{p}=\frac{1}{\sqrt{2}}\Big[(\dot{\rho}+i/\rho)\hat{a}^{+}+
(\dot{\rho}-i/\rho)\hat{a}\Big]~.
\end{equation}
Performing simple calculations, one gets
\begin{equation} \label{B3}
\hat{H}=f(\rho)\hat{a}^{+2}+f^{*}(\rho)\hat{a}^2+\frac{1}{4}\Big[\dot{\rho}^2+
\frac{1}{\rho ^2}+\omega ^2\rho ^2\Big](2\hat{I})~,
\end{equation}
where $f(\rho)=\dot{\rho} ^2+2i\dot{\rho}/\rho -1/\rho ^2+\omega ^2\rho ^2$.
Thus,
\begin{equation} \label{B4}
\langle n|\hat{H}|n\rangle=\langle n|\frac{1}{2}\Big[\dot{\rho}^2+
\frac{1}{\rho ^2}+\omega ^2\rho ^2\Big]\hat{I}|n\rangle
\end{equation}
from which eq.~(16) in chapter 5 is obvious.

\end{document}